\documentclass[10pt,twocolumn,english,prd,superscriptaddress,nofootinbib,preprintnumbers,showpacs,floatfix]{revtex4-2}

\usepackage{mathrsfs}
\usepackage{yfonts}
\usepackage{tipa}
\usepackage{bm}
\usepackage{bbm}
\usepackage{amsmath}
\usepackage{amssymb}
\usepackage{amsthm}
\usepackage{amsfonts}
\usepackage{mathtools}
\usepackage{latexsym}
\usepackage{relsize}
\usepackage[english]{babel}
\usepackage{booktabs}
\usepackage[iso-8859-7]{inputenc}
\usepackage{tikz}
\usepackage{xcolor}
\usepackage{array}
\usepackage{hyperref}
\usepackage{tabularx}
\usepackage{caption}

\newcommand{\bea}{\begin{eqnarray}}
	\newcommand{\eea}{\end{eqnarray}}
\newcommand{\be}{\begin{equation}}
	\newcommand{\ee}{\end{equation}}
\newcommand{\ba}{\begin{align}}
	\newcommand{\ea}{\end{align}}

\begin{document}
	
	\title{From St{\o}rmer to Schwarzschild:
		Analytical Dynamics of Charged Particles\\ in a Dipole Magnetosphere}
	
	\author{Francisco S. N. Lobo}
	\email{fslobo@ciencias.ulisboa.pt}
	\affiliation{Instituto de Astrof\'{i}sica e Ci\^{e}ncias do Espa\c{c}o,
		Faculdade de Ci\^{e}ncias da Universidade de Lisboa, Edificio C8,
		Campo Grande, P-1749-016 Lisbon, Portugal}
	\affiliation{Departamento de F\'{i}sica, Faculdade de Ci\^{e}ncias da
		Universidade de Lisboa, Edif\'{i}cio C8, Campo Grande,
		P-1749-016 Lisbon, Portugal}
	
	\author{Tiberiu Harko}
	\email{tiberiu.harko@aira.astro.ro}
	\affiliation{Department of Physics, Babe\c s-Bolyai University,
		Kog\u alniceanu Street, Cluj-Napoca 400084, Romania,}
	\affiliation{Astronomical Observatory, 19 Cire\c silor Street,
		400487 Cluj-Napoca, Romania}

\date{\today}

\begin{abstract}
We investigate the conservative dynamics of relativistic charged
particles in a Schwarzschild spacetime threaded by an externally
supported dipolar magnetic field. The gravitational sector is treated
exactly in isotropic coordinates, while the electromagnetic field is
described by the leading asymptotic term of the relativistic exterior
dipole solution. This formulation provides an analytically tractable
relativistic extension of the classical St{\o}rmer problem and allows
the circular-orbit and stability structure to be studied explicitly.
For the outward-Lorentz branch, characterized by a positive signed
magnetic coupling for positive azimuthal motion, we find a non-trivial
competition between radial and vertical stability. Weak magnetic
coupling shifts the radially marginal orbit inward, whereas at stronger
coupling vertical instability becomes the limiting mechanism and the
inner boundary of stable circular motion moves outward. The associated
vertical epicyclic mode progressively softens and vanishes at the
vertical stability boundary. The marginally bound sequence exhibits a
distinct strong-coupling behaviour, illustrating that energetic and
stability boundaries need not evolve together. The radial and vertical
epicyclic frequencies also generate characteristic commensurabilities
that may provide natural sites for nonlinear orbital coupling.
Because the magnetic field is treated as a test field, these
charged-particle effects leave the vacuum Schwarzschild null-geodesic
structure unchanged. Magnetic signatures in compact-object observations
must therefore arise through the dynamics and radiation of charged
matter, plasma propagation effects, or physics beyond the test-field
approximation. The framework developed here provides an analytical
basis for studying nonlinear dynamics, radiation reaction, and more
realistic compact-object magnetospheres.
\end{abstract}

\maketitle
	
\tableofcontents

\section{Introduction}
\label{sec:introduction}

The motion of charged particles in a dipole magnetic field, namely, the
St{\o}rmer problem, has played a fundamental role in space physics
since St{\o}rmer's pioneering studies of auroral particle
trajectories
\cite{St1,St2,St3,St4,St4a,St5,St6,St7}.
Its relativistic extension, the Classical Relativistic St{\o}rmer
Problem (CRSP), describes the Lorentz-force motion of a relativistic
charged particle in a prescribed dipolar field \cite{Jack,LL}.
Although symmetry allows important reductions of the dynamics,
generic trajectories can exhibit a rich nonlinear phase-space
structure, and the St{\o}rmer problem has continued to provide a
useful laboratory for studying particle trapping and orbital dynamics
in magnetic fields \cite{Harko:2026tev}. Charged-particle motion in
dipolar configurations has consequently been investigated in a wide
range of non-relativistic and relativistic settings
\cite{B1,B2,B3,Int,Dilao,VA1,VA2,Schust,How,Dull,In,In1,Epp0,Epp,Hal,
	Mark,Ozturk,Pina,Kol,Leg,Ersh,Asadi,Ersh1,Moc,Pap,Bur,Bur1}.

Around compact objects, however, the classical St{\o}rmer picture is
incomplete. Gravity can become comparable to, or compete directly
with, the electromagnetic force, while gravitational redshift,
spatial curvature, and the relativistic relation between conserved
and locally measured quantities modify the orbital dynamics. The
combined gravitational and Lorentz-force problem is therefore of
interest for understanding charged-particle trapping, acceleration,
and radiation in the environments of neutron stars and black holes.
It is also relevant indirectly to horizon-scale observations, since
magnetic fields strongly influence the dynamics and radiative
properties of the plasma responsible for the observed emission, for
example, by the Event Horizon Telescope
\cite{EventHorizonTelescope:2019dse}.

A particularly useful setting in which to investigate this interplay
is provided by a dipolar magnetic field on a Schwarzschild
background. Petterson~\cite{Petterson:1974bt} obtained an exact
exterior dipole solution of Maxwell's equations in Schwarzschild
spacetime. For a compact star, such an exterior field can be
supported by currents within the stellar interior. For a black hole,
by contrast, the decaying dipolar solution should be interpreted as
an externally supported test field, generated by currents in the
surrounding matter, rather than as an intrinsic magnetic dipole of
the Schwarzschild black hole. Charged-particle motion in related
Schwarzschild--dipole configurations has been investigated recently
by Stuchl\'{i}k et al.~\cite{Stuchlik:2024tlu}, including circular
motion and trapping regions. More broadly, relativistic
charged-particle dynamics in strong gravitational and electromagnetic
fields has been studied in a variety of compact-object contexts
\cite{Th,Thorne,Poisson2004,Preti2004,Lyutikov:2011vca,
	DiPiazza:2011tq,Cerutti:2015hvk,Philippov:2013tpa}.

The aim of the present work is to construct an analytically tractable
bridge between the relativistic St{\o}rmer problem and
charged-particle motion in Schwarzschild spacetime. We retain the
Schwarzschild geometry exactly in isotropic coordinates and represent
the electromagnetic sector by the leading asymptotic dipole term of
the relativistic exterior field. The model is therefore an exact
Schwarzschild gravitational problem coupled to a controlled
leading-dipole electromagnetic approximation, rather than the full
Schwarzschild--Petterson charged-particle system. Isotropic
coordinates are particularly convenient because the spatial metric
is conformally flat and provide a direct connection with the
Cartesian formulation of the classical St{\o}rmer problem.

Two distinctions are central to the analysis. First, the conserved
canonical angular momentum differs from the mechanical angular
momentum that determines the actual azimuthal motion. Second, the
signed coupling $\beta=e\mu/(mM^2)$ depends on both the electromagnetic
orientation and the particle specific charge. We concentrate on
positive mechanical azimuthal motion and on $\beta>0$, for which the
radial Lorentz force is outward. The complementary inward-Lorentz
branch $\beta<0$ has a qualitatively different stability structure
and is analysed comparatively in
Appendix~\ref{app:negative_beta}.

Circular-orbit stability must be examined in both radial and vertical
directions. Radial perturbations probe the curvature of the effective
potential, whereas vertical perturbations probe the angular structure
of the dipole field. These two conditions need not become marginal at
the same orbit, and their competition determines the inner boundary
of fully stable circular motion and the associated radial--vertical
frequency structure.

The photon sector remains conceptually separate. The coupling
$\beta$ applies only to massive charged particles, and the
electromagnetic field is treated as a test field that does not modify
the Schwarzschild metric. Photons therefore follow the usual
Schwarzschild null geodesics. Magnetic fields may nevertheless alter
observable emission through their influence on the plasma,
polarization, absorption, dispersion, and radiative transfer, without
changing the underlying vacuum photon sphere.

The astrophysical interpretation also requires care because $\beta$
is particle dependent and is not a property of the compact object
alone. Charged species or weakly charged structures in the same
external field can therefore occupy very different dynamical regimes,
and the stability boundary of a prescribed test charge should not be
identified directly with the inner edge of a quasi-neutral accretion
flow. The present conservative model is instead intended as a
reference problem for subsequent inclusion of radiation reaction,
nonlinear dynamics, and collective plasma effects; the dissipative
extension is pursued in the companion work~\cite{CRSP_react}.

The paper is organized as follows.
Section~\ref{sec:metric} introduces the Schwarzschild geometry and
dipolar field, and Sec.~\ref{sec:eom} develops the equations of motion
and effective potential. Section~\ref{sec:circular_catalogue}
analyses circular motion and its radial and vertical stability, while
Sec.~\ref{sec:vertical_epicyclic} develops the vertical dynamics and
frequency structure. Section~\ref{sec:additional_analytical}
presents complementary analytical results, and
Sec.~\ref{sec:applications} discusses the scope and astrophysical
interpretation of the model. We conclude in
Sec.~\ref{sec:conclusions}, while
Appendix~\ref{app:negative_beta} develops the complementary
$\beta<0$ branch and compares the two Lorentz-force orientations.
Except where factors of $G$ and $c$ are restored for physical
interpretation, geometric units $G=c=1$ are used.

\section{Metric and magnetic field}
\label{sec:metric}

\subsection{The Schwarzschild metric in isotropic Cartesian coordinates}
\label{sec:schwarzschild_isotropic}

We first fix the mass and radial-coordinate conventions used
throughout the paper. Let $M_\star$ denote the physical mass of the
central gravitating object and $M=GM_\star /c^2$
its geometrized mass length. The corresponding Schwarzschild radius
is $r_g=2M$. Thus, when factors of $G$ and $c$ are displayed
explicitly, $M$ has dimensions of length; in geometric units
$G=c=1$ it is the Schwarzschild mass parameter. The symbol $m$ is
reserved for the rest mass of the charged test particle.

In standard Schwarzschild coordinates $(t,r,\theta,\phi)$, with
$r$ the areal radius, the line element is \cite{LL}
\begin{align}
	ds^2
	&=
	\left(1-\frac{2M}{r}\right)c^2dt^2
	-\left(1-\frac{2M}{r}\right)^{-1}dr^2
	\nonumber\\
	&\quad
	-r^2
	\left(
	d\theta^2+\sin^2\theta\,d\phi^2
	\right).
	\label{eq:schwarzschild_areal}
\end{align}

Introducing the isotropic radial coordinate $R$ through
\begin{equation}
	r
	=
	R\left(1+\frac{M}{2R}\right)^2
	=
	R\left(1+\frac{r_g}{4R}\right)^2 ,
	\label{eq:areal_isotropic_relation}
\end{equation}
the metric becomes
\begin{equation}
	\begin{split}
		ds^2
		&=
		\left(
		\frac{1-\frac{M}{2R}}
		{1+\frac{M}{2R}}
		\right)^2 c^2dt^2 \\
		&\quad -
		\left(1+\frac{M}{2R}\right)^4
		\left(
		dR^2 + R^2d\theta^2 + R^2\sin^2\theta\,d\phi^2
		\right),
		\label{eq:schwarzschild_isotropic_spherical}
	\end{split}
\end{equation}
The Schwarzschild horizon $r=2M$ corresponds to $R=M/2=r_g/4$.
Throughout the charged-particle analysis we restrict attention to
the exterior isotropic region $R>M/2$.

To distinguish the spatial Cartesian coordinates from the
dimensionless isotropic radius $x=R/M$ introduced below, we denote
them by $(X,Y,Z)$. They are related to $(R,\theta,\phi)$ through
\begin{equation}
	X=R\sin\theta\cos\phi,
	\;\;\;
	Y=R\sin\theta\sin\phi,
	\;\;\;
	Z=R\cos\theta,
	\label{eq:isotropic_cartesian_transform}
\end{equation}
with
$R=\sqrt{X^2+Y^2+Z^2}$. The Schwarzschild metric then assumes the
spatially conformally flat form \cite{LL,Edd}
\begin{equation}
	\begin{split}
		ds^2
		&=
		\left(
		\frac{1-\dfrac{M}{2R}}
		{1+\dfrac{M}{2R}}
		\right)^2 c^2dt^2 \\
		&\quad -
		\left(1+\frac{M}{2R}\right)^4
		\left( dX^2+dY^2+dZ^2 \right),
		\label{metr}
	\end{split}
\end{equation}

For reference, the weak-field expansion of Eq.~\eqref{metr} is
\begin{align}
	&ds^2
	=
	\left[
	1-\frac{2M}{R}
	+\mathcal{O}\!\left(\frac{M^2}{R^2}\right)
	\right]c^2dt^2
	\nonumber\\	
	&\quad
	-
	\left[
	1+\frac{2M}{R}
	+\mathcal{O}\!\left(\frac{M^2}{R^2}\right)
	\right]
	\left(
	dX^2+dY^2+dZ^2
	\right).
	\label{eq:schwarzschild_weak_field}
\end{align}
Thus the metric approaches
$ds^2=c^2dt^2-dX^2-dY^2-dZ^2$ as $R\rightarrow\infty$, while its
leading corrections may equivalently be expressed in terms of
$r_g=2M$.

For the metric~\eqref{metr}, using $(t,X,Y,Z)$ as coordinates, the
determinant is
\begin{equation}
	g
	=
	-c^2
	\left(1-\frac{M}{2R}\right)^2
	\left(1+\frac{M}{2R}\right)^{10},
\end{equation}
and hence, in the exterior region,
\begin{equation}
	\sqrt{-g}
	=
	c
	\left(1-\frac{M}{2R}\right)
	\left(1+\frac{M}{2R}\right)^5 .
	\label{eq:sqrt_minus_g}
\end{equation}
If the temporal coordinate is instead chosen as $x^0=ct$, the
overall factor $c$ is absent from $\sqrt{-g}$.

The Cartesian coordinates $(X,Y,Z)$ may, when convenient, be
regarded as ordinary Cartesian coordinates of the auxiliary
Euclidean coordinate space associated with the isotropic chart.
This provides the algebraic identity
$R=\sqrt{X^2+Y^2+Z^2}$, but should not be confused with the physical
spatial geometry. On hypersurfaces of constant Schwarzschild time,
the physical spatial metric is conformally related to the Euclidean
one by the factor $\left(1+M/2R\right)^4$.

\subsection{The dipole magnetic field in isotropic Cartesian coordinates}
\label{sec:mag_field}

In curved spacetime, the electromagnetic field is described by the
four-potential one-form $A=A_\mu dx^\mu$ and the antisymmetric field
tensor
\begin{equation}
	F_{\mu\nu}
	=
	\partial_\mu A_\nu-\partial_\nu A_\mu .
	\label{eq:F_definition}
\end{equation}
The Maxwell equations are
\begin{align}
	\nabla_\lambda F_{\mu\nu}
	+\nabla_\nu F_{\lambda\mu}
	+\nabla_\mu F_{\nu\lambda}
	&=0,
	\label{M1}\\
	\nabla_\nu F^{\mu\nu}
	&=-\frac{4\pi}{c}j^\mu ,
	\label{M2}
\end{align}
where $j^\mu$ is the four-current. The sign in
Eq.~\eqref{M2} follows the electromagnetic convention adopted here.
Since the field considered below is evaluated in a vacuum region,
$j^\mu=0$, this sign does not affect the exterior solution.

\paragraph{The Schwarzschild dipole solution.}

We consider a static, axisymmetric magnetic dipole whose symmetry
axis coincides with the $Z$-axis. In Schwarzschild coordinates
$(t,r,\theta,\phi)$, the exterior vacuum dipole may be obtained from
the dipolar sector of the electromagnetic solution constructed by
Petterson~\cite{Petterson:1974bt}; see also
Ref.~\cite{Stuchlik:2024tlu}. The only non-vanishing covariant
component of the four-potential is
\begin{eqnarray}
	A_\phi(r,\theta)
	&=&
	-\frac{3\mu}{8M^3}\,
	r^2\sin^2\theta
	\Bigg[
	\ln\left(1-\frac{2M}{r}\right)
		\nonumber \\
	&&\qquad \quad+\frac{2M}{r}
	\left(1+\frac{M}{r}\right)
	\Bigg],
	\label{Aphi_exact}
\end{eqnarray}
where $\mu$ is the dipole parameter; its sign fixes the
orientation of the field relative to the chosen symmetry axis.

At large areal radius, Eq.~\eqref{Aphi_exact} becomes
\begin{equation}
	A_\phi(r,\theta)
	=
	\frac{\mu\sin^2\theta}{r}
	\left[
	1+\frac{3M}{2r}
	+\frac{12M^2}{5r^2}
	+\mathcal{O}\left(\frac{M^3}{r^3}\right)
	\right].
	\label{eq:Aphi_areal_expansion}
\end{equation}
The leading term therefore reproduces the standard asymptotic
dipolar dependence.

The decaying solution~\eqref{Aphi_exact} describes the exterior
field outside the currents or material source supporting the dipole.
It should not be interpreted as an intrinsic, horizon-regular
magnetic dipole of a Schwarzschild black hole: formal extrapolation
toward $r=2M$ encounters the logarithmic divergence in
Eq.~\eqref{Aphi_exact}. For a compact star the exterior solution is
applied outside the stellar surface and its supporting currents. For
a black hole the dipolar field must instead be regarded as externally
supported, for example by currents in the surrounding matter.
Throughout this work its stress-energy is assumed sufficiently small
that the Schwarzschild geometry remains fixed.

\paragraph{Transformation to isotropic coordinates.}

Using Eq.~\eqref{eq:areal_isotropic_relation}, the large-$R$
expansion of Eq.~\eqref{Aphi_exact} becomes
\begin{equation}
	A_\phi(R,\theta)
	=
	\frac{\mu\sin^2\theta}{R}
	\left[
	1+\frac{M}{2R}
	+\frac{3M^2}{20R^2}
	+\mathcal{O}\left(\frac{M^3}{R^3}\right)
	\right].
	\label{Aphi_expand}
\end{equation}
The coefficients differ from those in
Eq.~\eqref{eq:Aphi_areal_expansion} because the areal and isotropic
radii differ already at first order in $M/R$.

In isotropic Cartesian coordinates,
\[
d\phi
=
\frac{-Y\,dX+X\,dY}{X^2+Y^2}.
\]
Since $A=A_\phi d\phi$, the spatial covariant components of the
four-potential are
\begin{equation}
	A_X
	=
	-\frac{A_\phi Y}{X^2+Y^2},
	\quad
	A_Y
	=
	\frac{A_\phi X}{X^2+Y^2},
	\quad
	A_Z=0 .
	\label{eq:A_cov_cart}
\end{equation}

\paragraph{Leading dipole approximation.}

The analytical charged-particle model developed below retains the
leading term of Eq.~\eqref{Aphi_expand},
\[
A_\phi
=
\frac{\mu}{R}\sin^2\theta .
\]
The corresponding spatial potential one-form is
\begin{equation}
	A_i\,dX^i
	=
	\frac{\mu}{R^3}
	\left(
	-Y\,dX+X\,dY
	\right),
	\label{A_cart_leading}
\end{equation}
or, equivalently,
\begin{equation}
	(A_X,A_Y,A_Z)
	=
	\frac{\mu}{R^3}(-Y,X,0).
	\label{eq:A_cart_components}
\end{equation}

Equation~\eqref{eq:A_cart_components} gives the covariant spatial
components of the electromagnetic one-form. They should not be
identified directly with the components of a physical Euclidean
three-vector potential. Physical electric and magnetic fields are
observer-dependent quantities obtained by projecting
$F_{\mu\nu}$ onto an orthonormal frame, while raising spatial indices
in the coordinate basis introduces the inverse Schwarzschild metric
and the sign associated with the $(+---)$ signature.

For the particle dynamics, no separate three-vector construction is
required. From
$A_\phi=\mu\sin^2\theta/R$, Eq.~\eqref{eq:F_definition} gives
\begin{equation}
	F_{R\phi}
	=
	-\frac{\mu}{R^2}\sin^2\theta,
	\qquad
	F_{\theta\phi}
	=
	\frac{2\mu}{R}\sin\theta\cos\theta .
	\label{eq:F_dipole_leading}
\end{equation}
In the equatorial plane,
\begin{equation}
	F_{R\phi}
	=
	-\frac{\mu}{R^2},
	\qquad
	F_{\theta\phi}=0 .
	\label{eq:F_equatorial}
\end{equation}
The direction of the Lorentz force can therefore be determined
unambiguously from $F^\mu{}_\nu u^\nu$. In particular, the radial
force will be inferred below from $F^R{}_\phi u^\phi$, with the
required Schwarzschild metric factors entering automatically when the
index of $F_{R\phi}$ is raised.

The model thus combines the exact Schwarzschild gravitational
geometry with the leading asymptotic term of the relativistic dipole
potential. The first omitted electromagnetic correction is of
relative order $M/R$, so the approximation becomes systematically
accurate for $R/M\gg1$. At radii $R=O(M)$ the neglected curvature
corrections need not be small, and results there should be understood
as predictions of the leading-dipole analytical model rather than of
the full relativistic exterior field. Higher-order terms in
Eq.~\eqref{Aphi_expand}, or the full potential~\eqref{Aphi_exact}
within its appropriate domain of validity, provide the natural route
for quantifying these corrections.

\section{Charged-particle dynamics}
\label{sec:eom}

\subsection{Covariant equations of motion}
\label{sec:lagrangian}

The motion of a test particle of rest mass $m$ and electric charge
$e$ in prescribed gravitational and electromagnetic fields is
described by the action
\begin{equation}
	S
	=
	\int
	\left(
	-mc\,ds
	+\frac{e}{c}A_\mu dx^\mu
	\right),
	\label{action}
\end{equation}
where $ds=c\,d\tau$ along a timelike trajectory and $\tau$ denotes
the particle proper time. The particle is treated as a test body, so
its own gravitational and electromagnetic fields are neglected.

Equivalently, the same timelike trajectories follow from the
quadratic Lagrangian
\begin{equation}
	{\cal L}
	=
	\frac{m}{2}g_{\mu\nu}u^\mu u^\nu
	+
	\frac{e}{c}A_\mu u^\mu ,
	\label{Lagrangian_general}
\end{equation}
where $u^\mu=dx^\mu/d\tau$. The corresponding Euler--Lagrange
equations give the covariant Lorentz-force equation
\begin{equation}
	\frac{D u^\mu}{D\tau}
	=
	\frac{e}{mc}
	F^\mu{}_{\nu}u^\nu ,
	\label{Lorentz_covariant}
\end{equation}
where $F_{\mu\nu}$ is defined in Eq.~\eqref{eq:F_definition}. In
coordinate form,
\begin{equation}
	\frac{d^2x^\mu}{d\tau^2}
	+
	\Gamma^\mu_{\alpha\beta}
	\frac{dx^\alpha}{d\tau}
	\frac{dx^\beta}{d\tau}
	=
	\frac{e}{mc}
	F^\mu{}_{\nu}
	\frac{dx^\nu}{d\tau}.
	\label{Lorentz_explicit}
\end{equation}

For a massive particle parametrized by proper time, the four-velocity
satisfies
\begin{equation}
	g_{\mu\nu}u^\mu u^\nu=c^2 .
	\label{fourvelocity_norm}
\end{equation}
Together with the constants of motion associated with stationarity
and axial symmetry, this normalization condition provides the first
integral required below to construct the radial effective potential.

\subsection{Equatorial reduction and constants of motion}
\label{sec:constants}

The isotropic Cartesian formulation introduced in
Sec.~\ref{sec:schwarzschild_isotropic} provides a direct connection
with the classical St{\o}rmer problem. For the orbital analysis,
however, the axial symmetry of both the Schwarzschild geometry and
the aligned dipole field makes isotropic spherical coordinates more
convenient. We therefore specialize the covariant dynamics of
Sec.~\ref{sec:lagrangian} to the equatorial plane.

The plane $\theta=\pi/2$, corresponding to $Z=0$, is dynamically
invariant for initial conditions $\dot{\theta}=0$, since both the
geometry and the aligned dipole field are symmetric under reflection
across the equatorial plane. For the leading dipole potential
$A_\phi=\mu\sin^2\theta/R$, one has at $\theta=\pi/2$
\begin{equation}
	A_\phi=\frac{\mu}{R},
	\qquad
	F_{R\phi}=-\frac{\mu}{R^2},
	\qquad
	F_{\theta\phi}=0 ,
	\label{eq:equatorial_field_reduced}
\end{equation}
in agreement with Eq.~\eqref{eq:F_equatorial}.

Introducing
\begin{equation}
	\Phi(R)
	=
	\frac{1-M/(2R)}{1+M/(2R)},
	\qquad
	\Psi(R)
	=
	1+\frac{M}{2R},
	\label{eq:PhiPsi_dimensional}
\end{equation}
the Schwarzschild metric restricted to the equatorial plane becomes
\begin{equation}
	ds^2
	=
	\Phi^2(R)c^2dt^2
	-
	\Psi^4(R)
	\left(
	dR^2+R^2d\phi^2
	\right).
	\label{metric_equatorial}
\end{equation}

For the static magnetic configuration we choose a gauge with
$A_t=0$. The Lagrangian~\eqref{Lagrangian_general} then reduces to
\begin{align}
	{\cal L}
	&=
	\frac{m}{2}
	\left[
	\Phi^2c^2\dot{t}^{\,2}
	-
	\Psi^4
	\left(
	\dot{R}^{\,2}
	+R^2\dot{\phi}^{\,2}
	\right)
	\right]
	\nonumber\\
	&\quad
	+
	\frac{e}{c}A_\phi\dot{\phi}.
	\label{Lagrangian_equatorial}
\end{align}

Stationarity and axial symmetry make $t$ and $\phi$ cyclic
coordinates. The conserved momentum conjugate to $t$ is
\begin{equation}
	p_t
	=
	m\Phi^2c^2\dot{t}
	\equiv E .
\end{equation}
Defining the dimensionless specific energy
\begin{equation}
	{\cal E}
	=
	\frac{E}{mc^2},
\end{equation}
one obtains
\begin{equation}
	\dot{t}
	=
	\frac{{\cal E}}{\Phi^2}.
	\label{eq:tdot_constants}
\end{equation}

The canonical momentum conjugate to $\phi$ is
\begin{equation}
	p_\phi
	=
	-m\Psi^4R^2\dot{\phi}
	+
	\frac{e}{c}A_\phi .
	\label{eq:canonical_pphi}
\end{equation}
With the $(+---)$ signature, we define the conserved canonical axial
angular momentum by $L\equiv-p_\phi$, so that
\begin{equation}
	L
	=
	m\Psi^4R^2\dot{\phi}
	-
	\frac{e}{c}A_\phi .
	\label{Lconstant_sign}
\end{equation}

It is useful to distinguish the conserved canonical angular momentum
$L$ from the mechanical axial angular momentum,
\begin{equation}
	L_{\rm mech}
	\equiv
	m\Psi^4R^2\dot{\phi}
	=
	L+\frac{e}{c}A_\phi .
	\label{mechanical_L}
\end{equation}
Thus $L$ is conserved by axial symmetry, whereas $L_{\rm mech}$
determines the instantaneous azimuthal motion. The two coincide in
the neutral limit.

For the leading equatorial dipole, $A_\phi=\mu/R$, and therefore
Eq.~\eqref{mechanical_L} gives
\begin{equation}
	\dot{\phi}
	=
	\frac{1}{m\Psi^4R^2}
	\left(
	L+\frac{e\mu}{cR}
	\right).
	\label{eq:phidot_constants}
\end{equation}

The orientation of the Lorentz force follows directly from the
radial component of Eq.~\eqref{Lorentz_covariant}. Since
$g^{RR}=-\Psi^{-4}$ and $F_{R\phi}=-\mu/R^2$, raising the radial
index gives
\begin{equation}
	F^R{}_{\phi}
	=
	g^{RR}F_{R\phi}
	=
	\frac{\mu}{\Psi^4R^2}.
	\label{eq:FRphi_mixed}
\end{equation}
The electromagnetic contribution to the radial acceleration is
therefore
\begin{equation}
	\left(
	\frac{D u^R}{D\tau}
	\right)_{\rm em}
	=
	\frac{e\mu}{mc\,\Psi^4R^2}\,u^\phi .
	\label{eq:radial_Lorentz_sign}
\end{equation}

Since increasing $R$ defines the outward radial direction, positive
azimuthal motion $u^\phi>0$ experiences an outward Lorentz force
when $e\mu>0$ and an inward Lorentz force when $e\mu<0$. In terms
of the signed dimensionless coupling introduced below,
\begin{equation}
	\beta
	=
	\frac{e\mu}{mM^2},
\end{equation}
the branch $\beta>0$ therefore corresponds to the outward-Lorentz
orientation considered in the main analysis, whereas $\beta<0$
corresponds to the opposite orientation.

The conserved energy and canonical angular momentum, together with
Eqs.~\eqref{fourvelocity_norm} and \eqref{mechanical_L}, provide the
first integrals required to construct the radial effective potential.

\subsection{Effective potential for equatorial motion}
\label{sec:effective_potential}

The first integrals obtained in Sec.~\ref{sec:constants}, together
with the normalization condition~\eqref{fourvelocity_norm}, reduce
the equatorial radial dynamics to a one-dimensional
effective-potential problem. Substituting
$\dot{t}=\mathcal{E}/\Phi^2$ and
Eq.~\eqref{mechanical_L} into the normalization condition gives
\begin{equation}
	\frac{\Phi^2\Psi^4}{c^2}\,\dot{R}^{\,2}
	+
	V_{\rm eff}(R;L)
	=
	\mathcal{E}^2 ,
	\label{radial_eff}
\end{equation}
where
\begin{equation}
	V_{\rm eff}(R;L)
	=
	\Phi^2(R)
	\left[
	1+
	\frac{1}{m^2c^2\Psi^4(R)R^2}
	\left(
	L+\frac{e}{c}A_\phi
	\right)^2
	\right].
	\label{eq:Veff_dimensional}
\end{equation}
For the leading equatorial dipole, $A_\phi=\mu/R$, so the
electromagnetic interaction enters entirely through the
mechanical-angular-momentum combination
$L+e\mu/(cR)$.

Equation~\eqref{eq:Veff_dimensional} retains the exact Schwarzschild
metric factors together with the leading-dipole electromagnetic
interaction. The radial kinetic prefactor
$\Phi^2\Psi^4/c^2$ in Eq.~\eqref{radial_eff} is positive throughout
the exterior region $R>M/2$. It therefore does not alter the
location of turning points or the sign of radial stability, although
it must be retained when the curvature of the effective potential is
converted into a radial epicyclic frequency.

A circular orbit at $R=R_0$ has $\dot{R}=0$ and is an extremum of
$V_{\rm eff}$ at fixed canonical angular momentum $L$. Its energy
and equilibrium conditions are consequently
\begin{equation}
	\mathcal{E}^2
	=
	V_{\rm eff}(R_0;L),
	\qquad
	\left.
	\frac{\partial V_{\rm eff}}{\partial R}
	\right|_{R_0,L}
	=
	0 .
	\label{circular_conditions}
\end{equation}
Radial stability requires
\[
\left.
\frac{\partial^2 V_{\rm eff}}{\partial R^2}
\right|_{R_0,L}
>0 ,
\]
whereas a radially marginal circular orbit satisfies
\begin{equation}
	\left.
	\frac{\partial V_{\rm eff}}{\partial R}
	\right|_{R_0,L}
	=
	0,
	\qquad
	\left.
	\frac{\partial^2 V_{\rm eff}}{\partial R^2}
	\right|_{R_0,L}
	=
	0 .
	\label{ISCO_conditions}
\end{equation}
Equation~\eqref{ISCO_conditions} determines radial marginality only;
the inner boundary of fully stable circular motion must also satisfy
the independent vertical-stability condition developed below.

The effective potential~\eqref{eq:Veff_dimensional} applies to
massive charged particles and does not modify the null-geodesic
structure of the Schwarzschild spacetime. Within the test-field
approximation, the photon sphere therefore retains its standard
areal radius $r_{\rm ph}=3M$. In isotropic coordinates the same orbit
is located at
\begin{equation}
	R_{\rm ph}
	=
	M\left(1+\frac{\sqrt{3}}{2}\right)
	=
	\frac{M}{2}\left(2+\sqrt{3}\right).
	\label{Rph_correct}
\end{equation}
The photon sector is discussed separately in
Sec.~\ref{sec:photon_orbits}.

It is now convenient to introduce dimensionless isotropic variables,
which isolate the signed magnetic coupling and provide the natural
formulation for the circular-orbit and stability analysis.

\subsection{Dimensionless formulation}
\label{sec:dimensionless}

From this point onward we adopt geometric units $G=c=1$ and
introduce the dimensionless variables
\begin{equation}
	x\equiv\frac{R}{M},
	\qquad
	\lambda\equiv\frac{L}{mM},
	\qquad
	\beta\equiv\frac{e\mu}{mM^2}.
	\label{eq:dimless}
\end{equation}
We also define the dimensionless proper time $s=\tau/M$. Since
$R=Mx$, one has $dR/d\tau=dx/ds$.

The Schwarzschild metric functions become
\begin{equation}
	\Phi(x)
	=
	\frac{1-\dfrac{1}{2x}}
	{1+\dfrac{1}{2x}},
	\qquad
	\Psi(x)
	=
	1+\frac{1}{2x},
	\label{eq:PhiPsi_dimless}
\end{equation}
and the exterior isotropic region corresponds to $x>1/2$.

From Eq.~\eqref{mechanical_L}, the dimensionless mechanical axial
angular momentum is
\begin{equation}
	j
	\equiv
	\frac{L_{\rm mech}}{mM}
	=
	\lambda+\frac{\beta}{x}.
	\label{eq:j_general}
\end{equation}
For positive azimuthal motion, the main analysis considers
$\beta>0$, corresponding to the outward-Lorentz branch established
in Sec.~\ref{sec:constants}. The opposite Lorentz-force orientation,
$\beta<0$, is discussed separately in
Appendix~\ref{app:negative_beta}.

In these variables, the radial equation~\eqref{radial_eff} becomes
\begin{equation}
	\Phi^2(x)\Psi^4(x)
	\left(\frac{dx}{ds}\right)^2
	+
	V_{\rm eff}(x;\lambda,\beta)
	=
	\mathcal{E}^2 ,
	\label{eq:radial_dimless}
\end{equation}
where
\begin{equation}
	V_{\rm eff}(x;\lambda,\beta)
	=
	\Phi^2(x)
	\left[
	1+
	\frac{1}{\Psi^4(x)x^2}
	\left(
	\lambda+\frac{\beta}{x}
	\right)^2
	\right].
	\label{eq:Veff_dimless}
\end{equation}
Here $\mathcal{E}=E/m$. The Schwarzschild geometry is retained
exactly in Eq.~\eqref{eq:Veff_dimless}, while the electromagnetic
interaction is described by the leading dipole approximation
introduced in Sec.~\ref{sec:mag_field}.

\begin{figure}[th!]
	\centering
	\includegraphics[width=\columnwidth]{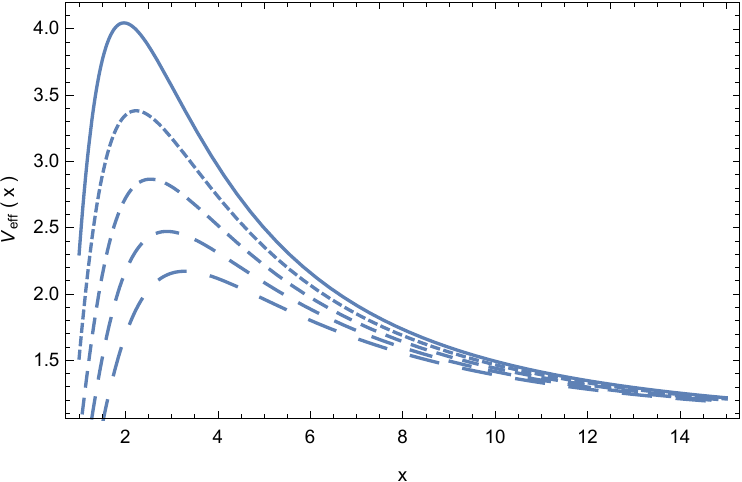}
	\caption{Effective potential $V_{\rm eff}(x)$ for
		$\lambda=10$ on the outward-Lorentz branch $\beta>0$, shown
		for representative values $\beta=0.01$, $2$, $4$, $6$, and
		$8$. The horizontal coordinate is the dimensionless
		isotropic radius $x=R/M$.}
	\label{fig1}
\end{figure}

For fixed canonical angular momentum, the magnetic coupling modifies
both the shape and the extrema of the effective potential, as
illustrated in Fig.~\ref{fig1}.

A circular orbit at $x=x_0$ satisfies
\begin{equation}
	V_{\rm eff}(x_0;\lambda,\beta)
	=
	\mathcal{E}^2,
	\qquad
	\left.
	\frac{\partial V_{\rm eff}}{\partial x}
	\right|_{x_0,\lambda,\beta}
	=
	0 .
	\label{eq:circ_cond}
\end{equation}
The derivative is taken at fixed canonical angular momentum
$\lambda$ and fixed magnetic coupling $\beta$.

The equilibrium condition in Eq.~\eqref{eq:circ_cond} is quadratic
in $\lambda$. Defining
\begin{equation}
	D(x)\equiv4x^2-8x+1 ,
\end{equation}
the two algebraic solutions are
\begin{equation}
	\begin{split}
		&\lambda_\pm(x,\beta)
		= \frac{1}{2xD(x)} \Bigl[  -\beta(12x^2-16x+1) \\
		&\quad \pm (2x+1)
		\sqrt{ \beta^2(2x-1)^2 + x(2x+1)^2D(x) } \Bigr],
	\end{split}
	\label{eq:lambda_circ}
\end{equation}
For the family considered in the main text, we select
$\lambda=\lambda_+(x,\beta)$, namely the branch continuously
connected to the positive-angular-momentum Schwarzschild
circular-orbit family as $\beta\rightarrow0$. The second algebraic
branch will be considered only when explicitly required.

In the neutral limit,
\begin{equation}
	\lambda_+(x,0)
	=
	\frac{(2x+1)^2}
	{2\sqrt{x(4x^2-8x+1)}} .
\end{equation}
The denominator vanishes when $D(x)=0$ at the exterior root
\begin{equation}
	x_{\rm ph}
	=
	1+\frac{\sqrt{3}}{2},
\end{equation}
which is the isotropic-coordinate location of the Schwarzschild
photon sphere given in Eq.~\eqref{Rph_correct}. The timelike
Schwarzschild circular-orbit sequence lies outside this limiting
radius.

Equations~\eqref{eq:Veff_dimless} and \eqref{eq:lambda_circ}
therefore determine the circular-orbit family and provide the basis
for the analysis of its conserved quantities, epicyclic frequencies,
and radial and vertical stability.

\section{Circular orbits in the Schwarzschild--dipole system}
\label{sec:circular_catalogue}

We now analyse equatorial circular trajectories of a charged test
particle in the Schwarzschild geometry with the external dipole field
introduced above. Throughout this section,
$x=R/M$ denotes the dimensionless isotropic radius. We consider
positive azimuthal motion on the outward-Lorentz branch $\beta>0$,
for which the dimensionless mechanical angular momentum is
$j=\lambda+\beta/x$.

Using the effective-potential formulation developed in
Sec.~\ref{sec:dimensionless}, we determine the conserved quantities
and characteristic frequencies along the Schwarzschild-connected
circular-orbit family and analyse its radial and vertical stability.
Since these two stability conditions are independent, the inner
boundary of fully stable circular motion is determined only after
both are taken into account.

\subsection{Equilibrium parameters of circular orbits}
\label{sec:equilibrium_params}

For a circular orbit at $x=x_0$, the canonical angular momentum on
the Schwarzschild-connected branch is
$\lambda_0=\lambda_+(x_0,\beta)$, where $\lambda_+$ is given by
Eq.~\eqref{eq:lambda_circ}. The corresponding dimensionless
mechanical angular momentum is
\begin{equation}
	j_0
	=
	\lambda_+(x_0,\beta)+\frac{\beta}{x_0}.
	\label{eq:j_circ}
\end{equation}
For $x_0>x_{\rm ph}$ and $\beta\geq0$, this branch satisfies
$j_0>0$ and therefore describes positive azimuthal motion.

The specific energy follows directly from
$V_{\rm eff}(x_0)=\mathcal{E}^2$:
\begin{equation}
	\mathcal{E}^2(x_0,\beta)
	=
	\Phi^2(x_0)
	\left[
	1+
	\frac{j_0^2}
	{\Psi^4(x_0)x_0^2}
	\right].
	\label{eq:E_circ}
\end{equation}
The positive square root is selected for future-directed motion in
the exterior region.

The azimuthal angular velocity with respect to Schwarzschild
coordinate time, $\Omega_\phi=d\phi/dt$, follows from the temporal
and azimuthal first integrals:
\begin{equation}
	M\Omega_\phi(x_0,\beta)
	=
	\frac{\Phi^2(x_0)}
	{\Psi^4(x_0)x_0^2}
	\frac{j_0}
	{\mathcal{E}(x_0,\beta)} .
	\label{eq:Omega_circ}
\end{equation}
Thus $M\Omega_\phi$ is the dimensionless orbital frequency measured
with respect to the asymptotic Schwarzschild time coordinate. It
should not be confused with a frequency defined with respect to
particle proper time or with that measured by a local static
observer.

In the neutral limit $\beta=0$, Eq.~\eqref{eq:Omega_circ} reduces to
\begin{equation}
	M\Omega_\phi
	=
	\frac{1}
	{x_0^{3/2}\Psi^3(x_0)}
	=
	\left(\frac{M}{r_0}\right)^{3/2},
	\label{eq:Omega_Schw_isotropic}
\end{equation}
where the corresponding areal radius is
$r_0=Mx_0\Psi^2(x_0)$. This is precisely the standard Schwarzschild
relation $\Omega_\phi^2=M/r_0^3$. Only in the asymptotic region
$x_0\gg1$, where $r_0\simeq Mx_0$, does it reduce to the approximate
isotropic-coordinate form $M\Omega_\phi\simeq x_0^{-3/2}$.

Equations~\eqref{eq:j_circ}--\eqref{eq:Omega_circ}, together with
the circular-orbit relation~\eqref{eq:lambda_circ}, determine the
canonical and mechanical angular momenta, specific energy, and
orbital frequency along the Schwarzschild-connected circular-orbit
family. These quantities define the equilibrium background about
which the radial and vertical perturbations are analysed below.

\subsection{Epicyclic frequencies}
\label{sec:epicyclic}

Small perturbations about a circular orbit probe its radial and
vertical stability. Since the radial kinetic term in
Eq.~\eqref{eq:radial_dimless} contains the position-dependent factor
$\Phi^2\Psi^4$, the curvature of the effective potential determines
the sign of radial stability but is not, by itself, the radial
epicyclic frequency.

Consider a radial perturbation $x=x_0+\delta x$, where $x_0$
satisfies the circular-orbit conditions~\eqref{eq:circ_cond}.
Linearizing the radial dynamics at fixed canonical angular momentum
$\lambda$ and fixed coupling $\beta$ gives
\begin{equation}
	\frac{d^2\delta x}{ds^2}
	+
	\omega_{r,s}^2\,\delta x
	=
	0,
	\qquad
	\omega_{r,s}^2
	=
	\frac{1}{2\Phi^2\Psi^4}
	\left.
	\frac{\partial^2 V_{\rm eff}}{\partial x^2}
	\right|_{x_0},
	\label{eq:omega_r}
\end{equation}
where $s=\tau/M$. Thus $\omega_{r,s}$ is the dimensionless
proper-time radial epicyclic frequency. Unless stated otherwise,
$\Phi$, $\Psi$, and the orbital quantities appearing below are
evaluated at $x=x_0$.

For convenience, we define the radial potential curvature
\begin{equation}
	{\cal K}_r(x;\lambda,\beta)
	\equiv
	\frac{1}{2}
	\frac{\partial^2 V_{\rm eff}}{\partial x^2},
	\label{eq:Kr_definition}
\end{equation}
where the derivative is taken at fixed $(\lambda,\beta)$. For the
effective potential~\eqref{eq:Veff_dimless}, one obtains
\begin{align}
	{\cal K}_r
	&=
	\frac{16}{(2x+1)^8}
	\Bigg\{
	8\beta^2\left[20(x-2)x+17\right]
	\nonumber\\
	&\quad
	+16\beta\lambda(3x-2)
	\left[4(x-2)x+1\right]
	\nonumber\\
	&\quad
	+\lambda^2
	\left[
	16x\{x[3(x-4)x+10]-2\}+1
	\right]
	\nonumber\\
	&\quad
	-2(x-1)(2x+1)^4
	\Bigg\}.
	\label{eq:omega_r_exact}
\end{align}
Hence
$\omega_{r,s}^2={\cal K}_r/(\Phi^2\Psi^4)$, with
${\cal K}_r$ evaluated on the circular-orbit family
$\lambda=\lambda_+(x_0,\beta)$.

For comparison with the orbital motion, it is useful to express the
epicyclic frequencies with respect to Schwarzschild coordinate time.
Since $d(t/M)/ds=\mathcal{E}/\Phi^2$, the dimensionless radial
frequency $\bar\Omega_r\equiv M\Omega_r$ is
\begin{equation}
	\bar\Omega_r^2
	=
	\frac{\Phi^4}{\mathcal{E}^2}\,
	\omega_{r,s}^2
	=
	\frac{\Phi^2}
	{2\mathcal{E}^2\Psi^4}
	\left.
	\frac{\partial^2 V_{\rm eff}}{\partial x^2}
	\right|_{x_0}.
	\label{eq:Omega_r_coordinate}
\end{equation}
The conversion factor between proper- and coordinate-time
frequencies is positive throughout the exterior region, so both
descriptions give the same radial stability criterion.

Vertical stability follows independently from a perturbation
$\theta=\pi/2+\zeta$ at fixed canonical axial angular momentum.
At linear order, the radial and vertical perturbations decouple.
The proper-time vertical frequency is
\begin{equation*}
	\omega_{\theta,s}^2
	=
	\frac{1}{\Psi^8 x_0^4}
	\left(
	j_0^2-\frac{2\beta j_0}{x_0}
	\right),
\end{equation*}
where
$j_0=\lambda_+(x_0,\beta)+\beta/x_0$.
A detailed derivation from the Euler--Lagrange equations is given in
Sec.~\ref{sec:Omega_z_derivation}, culminating in
Eq.~\eqref{eq:Omega_z}.

Converting to Schwarzschild coordinate time gives
\begin{equation*}
	\bar\Omega_\theta^2
	=
	\frac{\Phi^4}
	{\mathcal{E}^2\Psi^8 x_0^4}
	\left(
	j_0^2-\frac{2\beta j_0}{x_0}
	\right),
\end{equation*}
where $\bar\Omega_\theta\equiv M\Omega_\theta$. Together with
$\bar\Omega_\phi\equiv M\Omega_\phi$, the three characteristic
frequencies are therefore referred to the same time coordinate and
may be compared directly. In particular, as derived explicitly in
Eq.~\eqref{eq:vertical_orbital_ratio},
\begin{equation*}
	\frac{\Omega_\theta^2}{\Omega_\phi^2}
	=
	1-\frac{2\beta}{j_0x_0}.
\end{equation*}

For the outward-Lorentz branch $\beta>0$ with $j_0>0$, vertical
stability requires
\begin{equation*}
	j_0>\frac{2\beta}{x_0}.
\end{equation*}
Equivalently,
$\lambda_+(x_0,\beta)>\beta/x_0$. Whenever this condition is
satisfied, $0<\Omega_\theta<\Omega_\phi$. Thus the outward Lorentz
interaction reduces the vertical restoring frequency and may render
an orbit vertically unstable while it remains radially stable. On
the opposite Lorentz-force branch, discussed in
Appendix~\ref{app:negative_beta}, the magnetic contribution instead
enhances the vertical restoring frequency on the positive-$j$
family.

Radial marginal stability is characterized by
\begin{equation}
	\bar\Omega_r^2=0,
	\qquad\Longleftrightarrow\qquad
	\left.
	\frac{\partial^2 V_{\rm eff}}{\partial x^2}
	\right|_{x_0}
	=0,
	\label{eq:isco_epicyclic}
\end{equation}
together with the circular-orbit condition
$\partial_x V_{\rm eff}=0$. Vertical marginal stability is instead
defined by $\bar\Omega_\theta^2=0$, equivalently
$j_0=2\beta/x_0$. The inner boundary of the fully stable
circular-orbit sequence is therefore determined by whichever mode
first becomes marginal as the circular sequence is followed inward.

\begin{figure}[th!]
	\centering
	\includegraphics[width=\columnwidth]{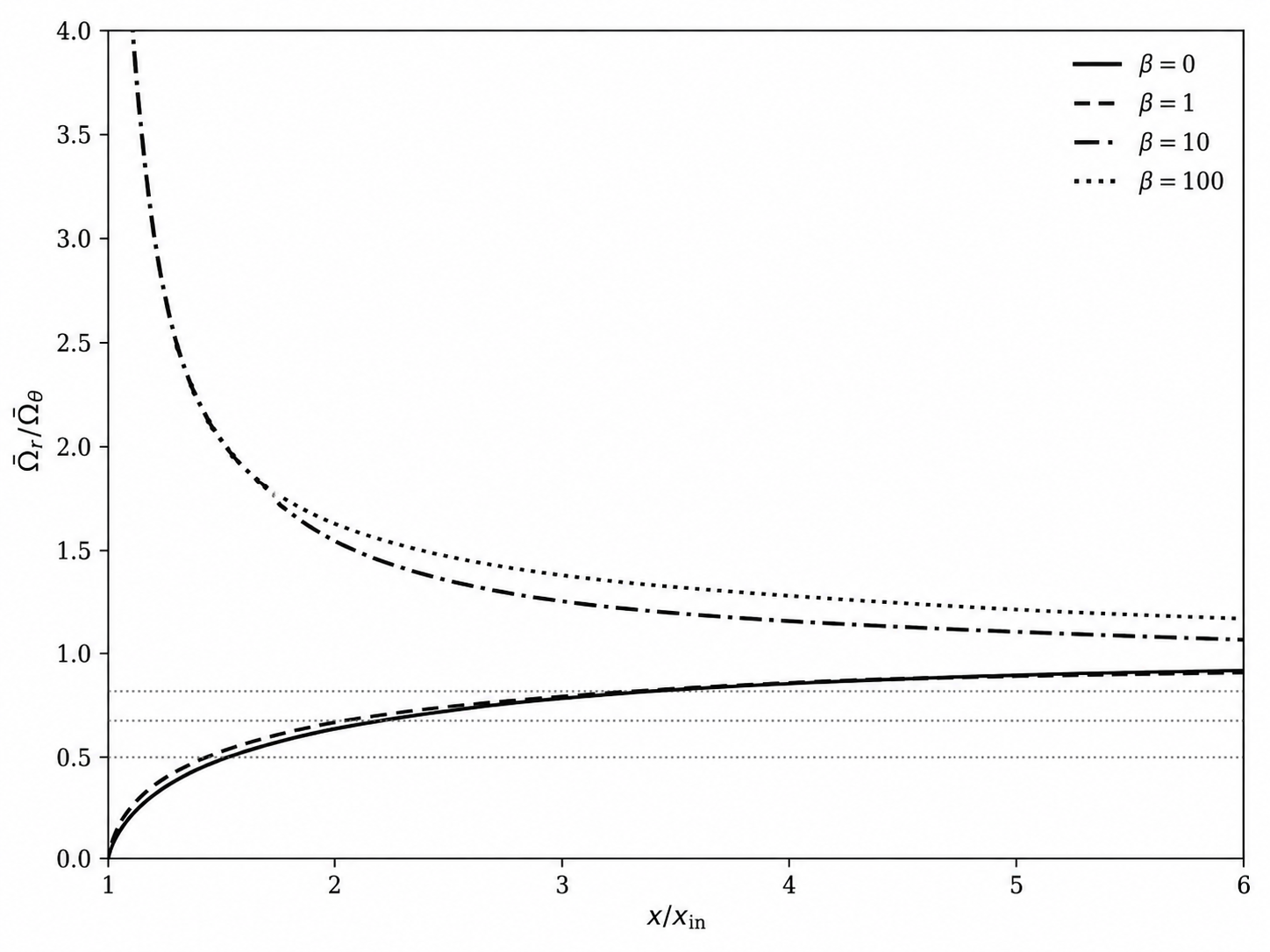}
	\caption{Ratio of the coordinate-time radial and vertical
		epicyclic frequencies,
		$\bar\Omega_r/\bar\Omega_\theta$, along the fully stable
		circular-orbit sequence for representative couplings
		$\beta=0$, $1$, $10$, and $100$. The isotropic radius is
		normalized by the corresponding inner stability boundary.
		Only regions satisfying
		$\bar\Omega_r^2>0$ and $\bar\Omega_\theta^2>0$ are shown.
		The horizontal dotted lines indicate representative
		sub-unity low-order commensurabilities.}
	\label{fig2}
\end{figure}

The ratio $\bar\Omega_r/\bar\Omega_\theta$ provides a useful
diagnostic of the relative behaviour of the two modes and of possible
low-order commensurabilities. Figure~\ref{fig2} shows this ratio
after imposing the circular-orbit relation
$\lambda=\lambda_+(x,\beta)$. This restriction is essential, since
evaluating the epicyclic expressions at arbitrary independent values
of $(x,\lambda,\beta)$ would not, in general, describe perturbations
about a circular background orbit.

The behaviour near the inner boundary depends on which mode becomes
marginal. At a radially marginal boundary,
$\bar\Omega_r\rightarrow0$ while $\bar\Omega_\theta$ remains finite,
so that
$\bar\Omega_r/\bar\Omega_\theta\rightarrow0$. At a vertically
marginal boundary,
$\bar\Omega_\theta\rightarrow0$ while $\bar\Omega_r$ remains finite,
and the ratio diverges. At sufficiently large radius the magnetic
corrections become subdominant and both modes recover their
Schwarzschild weak-field behaviour. The competition between these two
marginal-stability conditions determines the inner boundary of the
fully stable circular-orbit family, as analysed in the following
subsection.

\subsection{The ISCO as a function of the magnetic coupling}
\label{sec:isco_catalogue}

We now determine the inner boundary of the fully stable
circular-orbit sequence on the outward-Lorentz branch $\beta>0$.
As shown in Sec.~\ref{sec:epicyclic}, radial and vertical stability
must be imposed independently. Throughout this subsection, the term
ISCO refers to the innermost member of the \emph{fully stable}
circular-orbit sequence, irrespective of whether the limiting mode is
radial or vertical.

We begin with radial marginal stability. A radially marginal circular
orbit satisfies $\partial_xV_{\rm eff}=0$ and
$\partial_x^2V_{\rm eff}=0$ at fixed canonical angular momentum
$\lambda$ and fixed coupling $\beta$. Eliminating $\lambda$ gives
\begin{align}
	0={}&
	(2x-1)^2(2x+1)^4
	(4x^2-20x+1)^2
	\nonumber\\
	&-32\beta^2(2x+1)^2
	\left(
	80x^5-272x^4+368x^3
	\right.
	\nonumber\\
	&\hspace{2.4cm}
	\left.
	-224x^2+11x+1
	\right)
	\nonumber\\
	&-128\beta^4(2x-1)^2
	(4x^2-8x+7).
	\label{eq:isco_polynomial}
\end{align}
Although Eq.~\eqref{eq:isco_polynomial} contains only even powers of
$\beta$, its roots belong to different algebraic circular-orbit
branches. The radial marginal orbit relevant to the present analysis
must therefore also satisfy
$\lambda=\lambda_+(x,\beta)$ from
Eq.~\eqref{eq:lambda_circ} and be continuously connected to the
positive-angular-momentum Schwarzschild family.

In the neutral limit, the relevant exterior root is
\begin{equation}
	x_{\rm ISCO}^{(0)}
	=
	\frac{5+2\sqrt6}{2}
	\simeq4.94949 ,
	\label{eq:isco_schwarzschild_isotropic}
\end{equation}
which corresponds, through
$r/M=x(1+1/2x)^2$, to the standard Schwarzschild areal radius
$r_{\rm ISCO}=6M$.

For weak positive coupling, the radially marginal orbit behaves as
\begin{equation}
	x_{\rm rISCO}
	=
	x_{\rm ISCO}^{(0)}
	-
	\left(
	\frac{3\sqrt3}{8}
	+\frac{\sqrt2}{4}
	\right)\beta
	+
	\mathcal{O}(\beta^2).
	\label{eq:isco_small_b}
\end{equation}
The outward Lorentz interaction therefore initially shifts the radial
stability boundary inward.

Vertical marginal stability follows from
$j_0=2\beta/x$, equivalently
$\lambda=\beta/x$. Combining this condition with circularity gives
\begin{equation}
	\beta^2
	=
	\frac{x(2x+1)^4}
	{8(12x^2-16x+1)} .
	\label{eq:theta_isco_condition}
\end{equation}
This relation defines the vertically marginal circular-orbit curve.

The vertical marginal curve first reaches the boundary of the
exterior timelike circular-orbit domain at
$x=x_{\rm ph}=1+\sqrt3/2$, where
$\beta=\beta_{\theta,\min}
=\sqrt{9/2+21\sqrt3/8}\simeq3.00776$.
Its appearance does not immediately determine the inner boundary of
fully stable motion, because the vertically marginal orbit initially
lies inside the radially marginal one.

The radial and vertical marginal-stability curves intersect at
$x_{\rm c}\simeq2.05642$ and
$\beta_{\rm c}\simeq3.05318$, with $x_{\rm c}$ the relevant
exterior root of
$104x^3-236x^2+46x-1=0$. At this point both epicyclic modes become
marginal simultaneously, so
$\Omega_r=\Omega_\theta=0$.

The resulting stability structure is therefore piecewise. For
$0\leq\beta<\beta_{\rm c}$, the inner fully stable orbit is radially
marginal, $x_{\rm ISCO}=x_{\rm rISCO}$. In the narrow interval
$\beta_{\theta,\min}<\beta<\beta_{\rm c}$ a vertically marginal
circular orbit already exists, but it lies inside this radial
stability boundary and hence does not terminate the fully stable
sequence. At $\beta=\beta_{\rm c}$ the two marginal curves coincide.
For $\beta>\beta_{\rm c}$, vertical instability is encountered first
as the circular sequence is followed inward, and the ISCO is instead
the vertically marginal orbit.

The strong-coupling behaviour is therefore governed by
Eq.~\eqref{eq:theta_isco_condition}. Its large-radius expansion gives
$\beta^2\simeq x^3/6$, and hence
\begin{equation}
	x_{\rm ISCO}
	\sim
	6^{1/3}\beta^{2/3}
	\simeq
	1.81712\,\beta^{2/3},
	\qquad
	\beta\gg1 .
	\label{eq:isco_large_b}
\end{equation}
Thus the fully stable inner boundary eventually moves outward with
the characteristic $\beta^{2/3}$ scaling, but in this regime it is
set by vertical rather than radial marginality.

It is worth emphasizing that the eliminated radial equation
\eqref{eq:isco_polynomial} also possesses a large-radius asymptotic
root proportional to
$(5+3\sqrt3)^{1/3}\beta^{2/3}$. On the present $\beta>0$
orientation, however, this root does not lie on the
Schwarzschild-connected positive-$j$ circular-orbit family selected
by $\lambda_+(x,\beta)$ and therefore does not determine its physical
inner stability boundary. As shown in
Appendix~\ref{app:negative_beta}, the same asymptotic coefficient
becomes physically relevant on the opposite inward-Lorentz branch.

The dependence of the fully stable inner radius on the magnetic
coupling is consequently non-monotonic. Weak outward Lorentz support
moves the radially marginal boundary inward, whereas increasing the
coupling progressively weakens the vertical restoring force. Once
the radial and vertical marginal curves cross, the identity of the
limiting mode changes and the inner stability boundary reverses its
trend, moving outward at strong coupling. The large-$\beta$ regime
is therefore not a continuation of the weak-coupling radial ISCO
shift, but a qualitatively distinct vertically limited branch.

\subsection{Catalogue of circular orbits and analysis}
\label{sec:catalogue}

Table~\ref{tab:circular_catalogue} lists representative fully stable
circular orbits on the outward-Lorentz branch $\beta>0$, for
$\beta=0.1$, $1$, $10$, and $100$. All quantities are evaluated
from the isotropic-coordinate effective potential
\eqref{eq:Veff_dimless} after imposing the circular-orbit relation
$\lambda=\lambda_+(x,\beta)$. The radial coordinate is therefore
$x=R/M$ throughout.

For each orbit we give the specific energy $\mathcal{E}$, the
dimensionless canonical and mechanical angular momenta $\lambda$ and
$j$, and the coordinate-time frequencies $M\Omega_\phi$,
$M\Omega_r$, and $M\Omega_\theta$. Dimensional angular frequencies
are recovered by multiplying these quantities by
$c^3/(GM_\star)$, while the corresponding cyclic frequencies are
$\nu_i=c^3(M\Omega_i)/(2\pi GM_\star)$.

For the representative couplings in the table, the inner boundaries
of the fully stable circular sequence occur at
$x_{\rm ISCO}\simeq4.84891$, $3.92589$, $7.20946$, and $38.01821$
for $\beta=0.1$, $1$, $10$, and $100$, respectively. Here
$x_{\rm ISCO}$ denotes the innermost member of the \emph{fully
	stable} circular-orbit family, in the sense defined in
Sec.~\ref{sec:isco_catalogue}. The boundary is radially marginal for
the first two cases and vertically marginal for the latter two. The
first entry in each block of Table~\ref{tab:circular_catalogue} is
chosen slightly outside the corresponding stability boundary.

\begin{table*}[t]
	\centering
	\caption{Representative stable circular orbits on the
		$\beta>0$ outward-Lorentz branch. The coordinate
		$x=R/M$ is the dimensionless isotropic radius. All
		frequencies are defined with respect to Schwarzschild
		coordinate time and are reported in dimensionless form.}
	\label{tab:circular_catalogue}
	\begin{ruledtabular}
		\begin{tabular}{|cccccccc|}
			$\beta$ & $x$ & $\mathcal{E}$ & $\lambda$ & $j$
			& $M\Omega_\phi$ & $M\Omega_r$ & $M\Omega_\theta$ \\
			\hline
			0.1
			& 4.90 & 0.94167 & 3.42708 & 3.44749
			& 0.06863 & 0.00634 & 0.06823 \\
			& 6.00 & 0.94440 & 3.47371 & 3.49037
			& 0.05337 & 0.02151 & 0.05311 \\
			& 8.00 & 0.95239 & 3.65668 & 3.66918
			& 0.03677 & 0.02166 & 0.03665 \\
			& 10.00 & 0.95931 & 3.87564 & 3.88564
			& 0.02728 & 0.01859 & 0.02721 \\
			& 15.00 & 0.97069 & 4.42888 & 4.43554
			& 0.01559 & 0.01238 & 0.01556 \\
			\hline
			1
			& 4.00 & 0.92797 & 3.04450 & 3.29450
			& 0.08380 & 0.01050 & 0.07718 \\
			& 6.00 & 0.93868 & 3.21527 & 3.38194
			& 0.05202 & 0.02849 & 0.04939 \\
			& 8.00 & 0.94972 & 3.47078 & 3.59578
			& 0.03614 & 0.02426 & 0.03486 \\
			& 10.00 & 0.95781 & 3.73020 & 3.83020
			& 0.02693 & 0.01989 & 0.02622 \\
			& 15.00 & 0.97016 & 4.33451 & 4.40118
			& 0.01548 & 0.01276 & 0.01524 \\
			\hline
			10
			& 7.50 & 0.92212 & 1.50460 & 2.83793
			& 0.03236 & 0.04613 & 0.00795 \\
			& 10.00 & 0.94489 & 2.31967 & 3.31967
			& 0.02366 & 0.02969 & 0.01492 \\
			& 15.00 & 0.96525 & 3.40557 & 4.07223
			& 0.01439 & 0.01604 & 0.01180 \\
			& 20.00 & 0.97454 & 4.18133 & 4.68133
			& 0.00984 & 0.01042 & 0.00873 \\
			& 30.00 & 0.98333 & 5.35007 & 5.68340
			& 0.00562 & 0.00572 & 0.00528 \\
			\hline
			100
			& 40.00 & 0.98393 & 2.96499 & 5.46499
			& 0.00314 & 0.00528 & 0.00092 \\
			& 50.00 & 0.98787 & 4.39340 & 6.39340
			& 0.00239 & 0.00354 & 0.00146 \\
			& 60.00 & 0.99029 & 5.52758 & 7.19425
			& 0.00189 & 0.00257 & 0.00138 \\
			& 80.00 & 0.99307 & 7.30301 & 8.55301
			& 0.00128 & 0.00158 & 0.00108 \\
			& 100.00 & 0.99461 & 8.70523 & 9.70523
			& 0.00094 & 0.00109 & 0.00084 \\
		\end{tabular}
	\end{ruledtabular}
\end{table*}

Several qualitative trends are apparent.

\paragraph{Energy and angular momentum.}
At a fixed radius, increasing the outward magnetic coupling generally
reduces the mechanical angular momentum required for circular
equilibrium. The Lorentz force supplies part of the outward support
against gravity, thereby reducing the centrifugal contribution
required from the azimuthal motion. The canonical angular momentum
and specific energy change correspondingly. For every fixed finite
coupling, the magnetic correction becomes asymptotically weak and
the Schwarzschild large-radius behaviour is recovered, with
$\mathcal{E}\rightarrow1$ as $x\rightarrow\infty$.

\paragraph{Orbital frequency.}
At fixed radius, the reduction in mechanical angular momentum is
accompanied by a decrease in the coordinate-time orbital frequency
as the outward coupling is increased. The behaviour evaluated at the
inner stability boundary is different because the location and
character of that boundary themselves depend on $\beta$. At weak
coupling the radially marginal orbit moves inward, leading to a
larger orbital frequency at the inner edge. Once vertical marginal
stability becomes the limiting condition, the fully stable boundary
moves outward and its orbital frequency decreases. The evolution of
the characteristic inner-edge frequency is therefore non-monotonic.

\paragraph{Vertical and radial epicyclic motion.}
The vertical-frequency relation
\eqref{eq:vertical_orbital_ratio} shows that every vertically stable
orbit on the $\beta>0$ branch satisfies
$0<\Omega_\theta<\Omega_\phi$. The outward Lorentz interaction
therefore suppresses the vertical restoring frequency relative to
the orbital one. Equality is recovered in the neutral limit and
asymptotically at large radius, while at vertical marginal stability
the vertical frequency vanishes.

The radial mode is governed independently by the curvature of the
effective potential. When radial marginal stability determines the
inner boundary, $\Omega_r$ vanishes there while $\Omega_\theta$
remains finite. Once the sequence becomes vertically limited,
$\Omega_\theta$ instead vanishes at the inner boundary while
$\Omega_r$ remains finite. The identity of the soft mode therefore
changes across the transition identified in
Sec.~\ref{sec:isco_catalogue}.

\paragraph{Frequency ratios and commensurabilities.}
This distinction is reflected directly in
$\Omega_r/\Omega_\theta$. On a radially limited sequence the ratio
starts from zero at the inner boundary and increases outward through
sub-unity rational values. On a vertically limited sequence it
instead diverges as the inner boundary is approached from above and
then decreases through super-unity commensurabilities.

For the vertically limited branch, the ratio eventually crosses
unity at a finite radius. Beyond that crossing it becomes slightly
smaller than unity and approaches one from below in the asymptotic
weak-field region. The finite $1:1$ crossing is therefore distinct
from the large-radius degeneracy of the two epicyclic frequencies.
Depending on the coupling, the sub-unity excursion beyond the
crossing need not be large enough to encounter the same low-order
rational ratios characteristic of the weak-coupling sequences. The
detailed commensurability structure is examined in
Sec.~\ref{sec:resonances}.

Such rational crossings are kinematical properties of the linear
epicyclic spectrum. They identify possible locations for nonlinear
mode coupling, but do not by themselves establish resonance capture,
finite resonance widths, secular energy transfer, or chaotic
transport.

\paragraph{Non-monotonic inner stability boundary.}
The most important global feature of the catalogue is the
non-monotonic evolution of the fully stable inner orbit. Weak
outward-Lorentz coupling shifts the radially marginal boundary
inward. With increasing coupling, however, the vertical restoring
frequency is progressively suppressed until vertical marginal
stability becomes the limiting condition. The inner stable boundary
then reverses its trend and moves outward, asymptotically following
the strong-coupling behaviour derived in
Sec.~\ref{sec:isco_catalogue}.

This behaviour also illustrates the distinction between the
dimensionless particle coupling and the local strength of the
magnetic field. Although the dipolar interaction decreases with
radius, a sufficiently large $\beta$ can modify charged-particle
motion over an extended radial region. Conversely, a large value of
$\beta$ reflects the particle-specific charge-to-mass ratio as well
as the electromagnetic configuration and should not be interpreted
as a measure of field backreaction on the spacetime.

No independent Keplerian prescription should therefore be imposed on
the magnetically modified circular sequence. The orbital and
epicyclic timescales must be calculated from the actual charged-particle
frequencies derived above. The standard Schwarzschild--Kepler
behaviour is recovered only asymptotically, where the magnetic
contribution becomes subdominant.

\section{The vertical epicyclic frequency: derivation, analysis, and implications}
\label{sec:vertical_epicyclic}

The vertical epicyclic frequency characterizes small departures from
an equatorial circular orbit. Unlike the radial mode, which is
governed by the radial effective potential, the vertical mode probes
the angular structure of the dipole interaction and therefore
provides an independent stability condition. We derive it explicitly
here for the outward-Lorentz branch $\beta>0$.

\subsection{Derivation from the Euler--Lagrange equations}
\label{sec:Omega_z_derivation}

Consider an equatorial circular orbit at $x=x_0$ and introduce a
small latitudinal perturbation
$\theta=\pi/2+\zeta$, with $|\zeta|\ll1$. At linear order the
vertical perturbation decouples from the radial one, so the radius may
be held fixed at $x=x_0$. In geometric units, using the dimensionless
proper time $s=\tau/M$, the relevant angular part of the specific
Lagrangian is
\begin{equation}
	{\cal L}_{\rm ang}
	=
	-\frac{1}{2}\Psi^4x^2
	\left(
	\dot{\theta}^{\,2}
	+\sin^2\theta\,\dot{\phi}^{\,2}
	\right)
	+
	\frac{\beta}{x}\sin^2\theta\,\dot{\phi},
	\label{eq:L_vertical}
\end{equation}
where, in this subsection only, a dot denotes differentiation with
respect to $s$.

Conservation of the canonical axial angular momentum gives
\begin{equation}
	\Psi^4x^2\sin^2\theta\,\dot{\phi}
	=
	\lambda
	+
	\frac{\beta}{x}\sin^2\theta .
	\label{eq:phidot_off_equator}
\end{equation}
On the unperturbed equatorial orbit this becomes
$\Psi^4(x_0)x_0^2\dot{\phi}=j_0$, where
$j_0=\lambda_+(x_0,\beta)+\beta/x_0$.

The Euler--Lagrange equation for $\theta$ obtained from
Eq.~\eqref{eq:L_vertical} is
\begin{equation}
	\ddot{\theta}
	-\sin\theta\cos\theta\,\dot{\phi}^{\,2}
	+
	\frac{2\beta}{\Psi^4x^3}
	\sin\theta\cos\theta\,\dot{\phi}
	=0 .
\end{equation}
Using
$\sin(\pi/2+\zeta)=1+\mathcal{O}(\zeta^2)$ and
$\cos(\pi/2+\zeta)=-\zeta+\mathcal{O}(\zeta^3)$, and evaluating all
coefficients on the unperturbed circular background, one obtains
\begin{equation}
	\ddot{\zeta}
	+
	\omega_{\theta,s}^{\,2}\zeta
	=
	0,
	\label{eq:zeta_osc}
\end{equation}
where the dimensionless proper-time vertical epicyclic frequency is
\begin{equation}
	\omega_{\theta,s}^{\,2}
	=
	\frac{1}{\Psi^8(x_0)x_0^4}
	\left(
	j_0^2-\frac{2\beta j_0}{x_0}
	\right).
	\label{eq:Omega_z}
\end{equation}

The two terms in Eq.~\eqref{eq:Omega_z} have a transparent
interpretation. The first is the geometrical contribution associated
with azimuthal motion in the spherically symmetric Schwarzschild
background. In the neutral limit it reproduces the equality of the
vertical and orbital frequencies. The second originates from the
angular dependence of the dipole potential and gives the magnetic
correction to the vertical restoring force.

For the outward-Lorentz branch, $\beta>0$ and $j_0>0$, this magnetic
contribution is negative and therefore weakens vertical confinement.
Vertical stability requires
$\omega_{\theta,s}^2>0$, equivalently
$j_0>2\beta/x_0$. Since $j_0=\lambda+\beta/x_0$, the same condition
may be written as $\lambda>\beta/x_0$; equality defines vertical
marginal stability.

To compare the vertical mode with the radial and orbital frequencies,
we transform to Schwarzschild coordinate time using
$d(t/M)/ds=\mathcal{E}/\Phi^2$. The corresponding dimensionless
coordinate-time frequency,
$\bar\Omega_\theta\equiv M\Omega_\theta$, is
\begin{equation}
	\bar\Omega_\theta^2
	=
	\frac{\Phi^4(x_0)}
	{\mathcal{E}^2\Psi^8(x_0)x_0^4}
	\left(
	j_0^2-\frac{2\beta j_0}{x_0}
	\right).
	\label{eq:Omega_theta_coord_vertical}
\end{equation}
Combining this result with the orbital frequency
Eq.~\eqref{eq:Omega_circ} gives
\begin{equation}
	\frac{\Omega_\theta^2}{\Omega_\phi^2}
	=
	1-\frac{2\beta}{j_0x_0}.
	\label{eq:vertical_orbital_ratio}
\end{equation}

Equation~\eqref{eq:vertical_orbital_ratio} immediately recovers
$\Omega_\theta=\Omega_\phi$ in the Schwarzschild limit
$\beta=0$. On every vertically stable orbit of the $\beta>0$ branch
one instead has $0<\Omega_\theta<\Omega_\phi$. At vertical marginal
stability, $j_0=2\beta/x_0$, the vertical mode softens completely and
$\Omega_\theta=0$.

The physical effect of the outward Lorentz interaction is therefore
particularly clear: increasing the magnetic coupling reduces the
vertical restoring force and introduces a stability mechanism
independent of radial marginality. For sufficiently strong coupling,
vertical instability can consequently terminate the fully stable
circular-orbit sequence before radial instability is reached, as
discussed in Sec.~\ref{sec:isco_catalogue}. For the opposite
Lorentz-force orientation, considered in
Appendix~\ref{app:negative_beta}, the magnetic contribution changes
sign and instead enhances the vertical restoring frequency on the
positive-$j$ circular-orbit family.

\subsection{Limiting cases, parametric dependence, and physical interpretation}
\label{sec:Omega_z_limits}

The behaviour of the vertical epicyclic frequency is particularly
transparent in three regimes: the Schwarzschild limit, the
large-radius limit at fixed magnetic coupling, and the
strong-coupling region near the inner stability boundary. Throughout
this subsection we consider the outward-Lorentz branch $\beta>0$.

\subsubsection{Schwarzschild limit ($\beta\rightarrow0$)}

In the neutral limit the canonical and mechanical angular momenta
coincide, $j_0=\lambda$, and
Eq.~\eqref{eq:vertical_orbital_ratio} immediately gives
$\Omega_\theta=\Omega_\phi$, provided that both frequencies are
referred to the same Schwarzschild coordinate time. Using
Eq.~\eqref{eq:Omega_Schw_isotropic}, their common value is
$M\Omega_\theta=M\Omega_\phi
=x^{-3/2}\Psi^{-3}(x)=(M/r)^{3/2}$.

This equality is a consequence of spherical symmetry. A slightly
inclined Schwarzschild circular geodesic lies in a neighbouring
orbital plane, so the vertical coordinate oscillation has the same
coordinate-time frequency as the azimuthal motion.

\subsubsection{Large-radius limit at fixed magnetic coupling}

For fixed $\beta$ and $x\rightarrow\infty$, the dipolar interaction
becomes asymptotically weak and the circular-orbit sequence approaches
the Schwarzschild weak-field family. In particular,
$j_0\sim x^{1/2}$, $\mathcal{E}\rightarrow1$, and
$M\Omega_\phi\sim x^{-3/2}$. Equation
\eqref{eq:vertical_orbital_ratio} then gives
$\Omega_\theta^2/\Omega_\phi^2
=1-2\beta/(j_0x)
=1-\mathcal{O}(\beta x^{-3/2})$.
The magnetic suppression of the vertical frequency therefore
disappears asymptotically, with
$\Omega_\theta/\Omega_\phi\rightarrow1$ from below.

The radial frequency has the same leading weak-field scaling, so
$M\Omega_\phi\sim M\Omega_r\sim M\Omega_\theta\sim x^{-3/2}$ at
sufficiently large radius.

This limit should not be confused with a magnetically supported
circular orbit in flat spacetime. On the branch considered here,
positive azimuthal motion experiences an outward Lorentz force, so
both the centrifugal and electromagnetic contributions act outward
and gravity remains essential for radial equilibrium. A non-trivial
magnetically confined circular orbit in the auxiliary gravity-free
St{\o}rmer problem requires the opposite Lorentz-force orientation,
as discussed in Appendix~\ref{app:negative_beta}.

\subsubsection{Strong-coupling behaviour near the ISCO}

For $\beta>\beta_{\rm c}$, the inner boundary of the fully stable
circular-orbit sequence is determined by vertical marginal stability.
In the strong-coupling regime,
Eq.~\eqref{eq:theta_isco_condition} gives
\begin{equation}
	x_{\rm ISCO}
	\sim
	C_\theta\,\beta^{2/3},
	\qquad
	C_\theta\equiv6^{1/3}
	\simeq1.81712 .
	\label{eq:x_isco_strong_vertical}
\end{equation}
Thus $x_{\rm ISCO}\gg1$ as $\beta\rightarrow\infty$, so gravitational
redshift and spatial-curvature corrections become progressively
small. Gravity nevertheless remains dynamically essential: on this
scaling its inward attraction continues to balance the centrifugal
and outward electromagnetic contributions.

Vertical marginal stability implies
$j_{\rm ISCO}=2\beta/x_{\rm ISCO}$, and hence
$j_{\rm ISCO}\sim2\,6^{-1/3}\beta^{1/3}$. Since
$\Phi\rightarrow1$, $\Psi\rightarrow1$, and
$\mathcal{E}\rightarrow1$ along this large-radius sequence,
Eq.~\eqref{eq:Omega_circ} then gives
$M\Omega_{\phi,\rm ISCO}\sim(3\beta)^{-1}$.

The qualitative distinction from a radially marginal inner orbit is
particularly clear in the vertical mode. Once vertical marginal
stability determines the inner boundary,
$\Omega_{\theta,\rm ISCO}=0$ identically, rather than merely in the
large-$\beta$ limit. The radial epicyclic frequency remains finite
there. The strong-coupling inner orbit is therefore characterized by
a soft vertical mode rather than by radial marginality.

\subsubsection{Systematic trends}

Several general features follow directly from these limits. Every
vertically stable circular orbit on the $\beta>0$ branch satisfies
$0<\Omega_\theta<\Omega_\phi$, so the outward Lorentz interaction
reduces the vertical restoring frequency relative to the orbital
frequency. Vertical stability is lost when
$j_0=2\beta/x_0$, at which point $\Omega_\theta$ vanishes.

At fixed coupling, the magnetic splitting of the characteristic
frequencies decreases with increasing radius and the Schwarzschild
weak-field behaviour is recovered asymptotically. Along the inner
stability boundary, however, the dependence on $\beta$ changes
qualitatively. At weak coupling radial marginal stability is the
limiting condition and the inner stable orbit moves inward. At
sufficiently strong coupling vertical marginal stability becomes
controlling and the inner boundary instead moves outward according to
Eq.~\eqref{eq:x_isco_strong_vertical}.

The two regimes are therefore distinguished by different soft modes:
$\Omega_r$ vanishes at a radially marginal boundary, whereas
$\Omega_\theta$ vanishes at a vertically marginal one. This change is
central to the behaviour of the radial--vertical frequency ratio and
to the pattern of commensurabilities along the stable circular-orbit
sequence.

These conclusions are specific to the outward-Lorentz orientation.
For the complementary $\beta<0$, positive-$j$ branch analysed in
Appendix~\ref{app:negative_beta}, the magnetic contribution instead
enhances the vertical restoring frequency and the inner stability
boundary remains radially determined.

\subsection{Commensurabilities between the radial and vertical motions}
\label{sec:resonances}

The radial and vertical epicyclic frequencies provide a natural
diagnostic of locations at which the two linear oscillation modes
become commensurate. On the outward-Lorentz branch $\beta>0$, the
behaviour of the ratio $\Omega_r/\Omega_\theta$ depends qualitatively
on the mechanism that determines the inner boundary of fully stable
circular motion.

When radial marginal stability sets the inner boundary,
$\Omega_r\rightarrow0$ while $\Omega_\theta$ remains finite as
$x\rightarrow x_{\rm ISCO}^{+}$, and hence
$\Omega_r/\Omega_\theta\rightarrow0$. When vertical marginal
stability instead sets the boundary,
$\Omega_\theta\rightarrow0$ while $\Omega_r$ remains finite, so that
$\Omega_r/\Omega_\theta\rightarrow\infty$. At sufficiently large
radius and fixed $\beta$, the magnetic corrections become
subdominant and the ratio approaches unity as the Schwarzschild
weak-field regime is recovered.

A relation $\Omega_r/\Omega_\theta=p/q$, with integers $p$ and $q$,
defines a commensurability between the two linearized modes. Such a
condition identifies a candidate location for nonlinear mode
coupling, but does not by itself establish a dynamical resonance.
Resonance widths, capture probabilities, secular energy exchange,
and possible chaotic transport depend on nonlinear terms that are
absent from the linear epicyclic analysis.

Table~\ref{tab:resonances} lists representative low-order
commensurabilities on the outward-Lorentz circular-orbit branch. The
radii are obtained from the coordinate-time frequencies after
imposing $\lambda=\lambda_+(x,\beta)$ at every point and restricting
to the region in which both epicyclic modes are stable. A dash
indicates that the specified ratio is not attained on the stable
sequence for the coupling shown.

\begin{table}[t]
	\centering
	\caption{Isotropic radii $x_{\rm res}=R_{\rm res}/M$ at which
		selected low-order commensurabilities
		$\Omega_r:\Omega_\theta=p:q$ occur on the stable
		$\beta>0$ circular-orbit branch. For the weak-coupling
		examples the $1:1$ ratio is reached only asymptotically,
		whereas for the strong-coupling examples it is crossed at a
		finite radius.}
	\label{tab:resonances}
	\begin{ruledtabular}
		\begin{tabular}{|c|c|c|c|c|}
			$\Omega_r:\Omega_\theta$
			& $\beta=0.1$ & $\beta=1$ & $\beta=10$ & $\beta=100$ \\
			\hline
			$3:1$ & -- & -- & 8.338 & 45.576 \\
\hline
			$2:1$ & -- & -- & 9.969 & 56.508 \\
\hline
			$3:2$ & -- & -- & 12.965 & 77.180 \\
\hline
			$4:3$ & -- & -- & 15.521 & 95.631 \\
\hline
			$5:4$ & -- & -- & 17.745 & 112.388 \\
\hline
			$1:1$ & $\infty$ & $\infty$ & 97.289 & 9997.333 \\
\hline
			$4:5$ & 15.326 & 12.087 & -- & -- \\
\hline
			$3:4$ & 12.417 & 9.656 & -- & -- \\
\hline
			$2:3$ & 9.552 & 7.356 & -- & -- \\
\hline
			$1:2$ & 6.805 & 5.287 & -- & -- \\
\hline
			$1:3$ & 5.582 & 4.424 & -- & -- \\
		\end{tabular}
	\end{ruledtabular}
\end{table}

Several qualitative features follow from
Table~\ref{tab:resonances} and Fig.~\ref{fig2}.

\paragraph{Radially limited regime.}
For $\beta<\beta_{\rm c}$, the frequency ratio starts from zero at
the radially marginal inner boundary and increases outward through
sub-unity commensurabilities. The representative cases
$\beta=0.1$ and $1$ illustrate how these crossings shift as the
outward Lorentz interaction modifies the circular-orbit sequence. In
these examples the ratio remains below unity and reaches $1:1$ only
asymptotically.

\paragraph{Vertically limited regime.}
For $\beta>\beta_{\rm c}$, the behaviour near the inner boundary is
qualitatively different. Since the vertical mode softens while the
radial frequency remains finite,
$\Omega_r/\Omega_\theta\rightarrow\infty$ as
$x\rightarrow x_{\rm ISCO}^{+}$. Moving outward, the ratio decreases
through super-unity commensurabilities such as $3:1$, $2:1$, $3:2$,
$4:3$, and $5:4$. This pattern is a direct consequence of vertical,
rather than radial, marginality at the inner stability boundary.

\paragraph{The $1:1$ crossing and the asymptotic regime.}
On the vertically limited branch, the decreasing ratio crosses unity
at a finite radius. Beyond this point it lies below unity and
eventually approaches $\Omega_r/\Omega_\theta=1$ from below as the
weak-field regime is recovered. The finite $1:1$ crossing is
therefore distinct from the asymptotic degeneracy of the radial and
vertical frequencies in the Newtonian limit.

The large radius of this crossing at strong coupling reflects the
competition between corrections with different radial dependences.
Eventually the magnetic contribution to the frequency splitting
becomes subdominant and the familiar Schwarzschild weak-field
ordering is recovered, even though the intermediate magnetically
modified region may extend over a large radial range.

\paragraph{Nonlinear interpretation.}
The commensurabilities identified above are kinematical properties of
the linear epicyclic spectrum. Whether they generate dynamically
important resonances depends on nonlinear coupling between the radial
and vertical degrees of freedom. Determining resonance widths,
capture probabilities, secular energy transfer, resonance overlap,
or chaotic transport therefore requires either a higher-order
perturbative treatment or direct numerical exploration of the phase
space.

The relativistic St{\o}rmer problem provides a strongly nonlinear
setting in which numerical trajectories complement the analytical
description~\cite{Harko:2026tev}. In the classical St{\o}rmer
problem, trapped trajectories are known to display periodic,
quasiperiodic, and chaotic behaviour~\cite{Dilao}. Whether analogous
nonlinear structures are organized by the radial--vertical
commensurabilities identified here is an interesting question for the
Schwarzschild--dipole system, but lies beyond the linear stability
analysis of the present work.

The complementary inward-Lorentz orientation has a qualitatively
different epicyclic hierarchy. As shown in
Appendix~\ref{app:negative_beta}, its positive-$j$ circular-orbit
family remains radially limited and exhibits enhanced vertical
restoring frequencies, without the vertically soft inner boundary
characteristic of the strong-coupling $\beta>0$ branch.

\subsection{Implications for trapped particle populations}
\label{sec:implications_vertical}

The vertical epicyclic frequency characterizes the response of an
individual charged particle to small departures from a stable
equatorial circular orbit. It therefore provides a measure of local
vertical confinement and a characteristic dynamical timescale for
dilute trapped-particle populations. Extending these results to a
collective plasma, however, requires additional physics beyond the
test-particle approximation.

\subsubsection{Vertical excursions and geometrical thickness}

For the linearized motion described by
Eq.~\eqref{eq:zeta_osc}, the dimensionless vertical oscillator energy
is
\begin{equation}
	{\cal E}_\theta
	=
	\frac{1}{2}\dot{\zeta}^{\,2}
	+
	\frac{1}{2}\omega_{\theta,s}^2\zeta^2 ,
	\label{eq:Etheta}
\end{equation}
where the dot denotes differentiation with respect to the
dimensionless proper time $s=\tau/M$. For a vertically stable orbit,
$\omega_{\theta,s}^2>0$, the maximum angular displacement is
\begin{equation}
	|\zeta|_{\rm max}
	=
	\frac{\sqrt{2{\cal E}_\theta}}
	{\omega_{\theta,s}} .
	\label{eq:zeta_max}
\end{equation}

Since $\zeta$ is an angular variable, the corresponding local proper
meridional excursion at fixed $R$ is, for $|\zeta|\ll1$,
\begin{equation}
	H_{\rm p}
	\simeq
	\Psi^2(R)\,R\,|\zeta|_{\rm max}
	=
	M\Psi^2(x)x\,
	\frac{\sqrt{2{\cal E}_\theta}}
	{\omega_{\theta,s}} .
	\label{eq:vertical_excursion}
\end{equation}
Thus, for a prescribed oscillator energy on a given circular
background, a smaller vertical epicyclic frequency corresponds to
weaker confinement and a larger vertical excursion.

On the outward-Lorentz branch $\beta>0$, the dipole interaction
reduces the vertical restoring frequency. As vertical marginal
stability is approached,
$\omega_{\theta,s}\rightarrow0$, and
Eq.~\eqref{eq:vertical_excursion} formally predicts an unbounded
excursion at fixed ${\cal E}_\theta$. This divergence is not physical:
it indicates the breakdown of the harmonic approximation once the
predicted amplitude is no longer small. Higher-order terms in the
latitudinal dynamics must then be retained.

Equation~\eqref{eq:vertical_excursion} characterizes individual
particles, or a dilute ensemble with a specified distribution of
vertical oscillator energies. It should not be interpreted as the
hydrostatic scale height of an accretion flow, which depends on
pressure, magnetic stresses, currents, radiation, and collective
plasma dynamics.

At large radius and fixed $\beta$,
$\omega_{\theta,s}\propto x^{-3/2}$. Comparing circular orbits at
fixed ${\cal E}_\theta$ therefore gives
$H_{\rm p}\propto x^{5/2}$, up to the asymptotically negligible
factor $\Psi^2\rightarrow1$. This scaling follows specifically from
holding the dimensionless oscillator energy fixed and is not a
universal prediction for the thickness or radial evolution of a
particle population.

\subsubsection{Characteristic variability timescales}

Vertical motion also introduces a characteristic coordinate-time
scale. The corresponding physical cyclic frequency is
\begin{equation}
	\nu_\theta
	=
	\frac{c^3}{2\pi G M_\star}\,
	\bar\Omega_\theta
	=
	\frac{c}{2\pi M}\,
	\bar\Omega_\theta ,
	\label{eq:nu_theta_physical}
\end{equation}
where $\bar\Omega_\theta=M\Omega_\theta$ and
$M=GM_\star/c^2$. At the inner boundary of the fully stable
circular-orbit sequence,
\begin{equation}
	\nu_\theta^{\rm ISCO}
	=
	\frac{c^3}{2\pi G M_\star}\,
	\bar\Omega_\theta(x_{\rm ISCO}) .
	\label{eq:nu_z_isco}
\end{equation}
The dimensional scale therefore varies as $M_\star^{-1}$, while the
dimensionless frequency reflects the magnetic coupling and the
stability mechanism that determines the inner orbit.

In the Schwarzschild limit, the vertical and orbital frequencies
coincide at the ISCO. On the $\beta>0$ branch, the vertical mode is
progressively softened by the magnetic interaction. When radial
marginality determines the inner boundary, $\Omega_\theta$ remains
finite there. At the transition between the radial and vertical
stability regimes the vertical frequency vanishes together with the
radial one, whereas once vertical marginality controls the inner
boundary,
\begin{equation}
	\bar\Omega_\theta(x_{\rm ISCO})=0,
	\qquad
	\beta>\beta_{\rm c}.
\end{equation}
The associated linear vertical oscillation time then formally
diverges. This should be interpreted as the emergence of a soft mode,
not as an observable oscillation of infinite period.

Any connection between $\nu_\theta$ and an observed periodic or
quasi-periodic signal requires a model for excitation and emission.
The modulation can depend on the particle distribution, viewing
geometry, and radiative-transfer effects. Moreover, because the
aligned configuration is symmetric under
$\zeta\rightarrow-\zeta$, an observable that is itself even under
this transformation need not contain the fundamental vertical
frequency. Its leading oscillatory contribution may instead occur at
$2\Omega_\theta$, with higher even harmonics generated by nonlinear
dependence on the vertical displacement.

\section{Additional analytical results}
\label{sec:additional_analytical}

The preceding analysis determines the circular-orbit family and its
radial and vertical stability within the isotropic-coordinate
leading-dipole model. We now consider two complementary orbital
scales: the marginally bound circular orbit of the charged-particle
system and, separately, the circular null geodesic of the underlying
Schwarzschild geometry. The latter provides a useful reference for
distinguishing Lorentz-force effects on massive charged particles
from the vacuum null-geodesic structure of the fixed spacetime.

\subsection{Marginally bound circular orbits}
\label{sec:marginally_bound}

A marginally bound circular orbit has specific energy
$\mathcal{E}=1$. Since $V_{\rm eff}\rightarrow1$ as
$x\rightarrow\infty$, this is the rest-energy threshold for a
particle at infinity.

On the outward-Lorentz branch $\beta>0$, the marginally bound radius
$x_{\rm mb}$ is determined by
\begin{equation}
	V_{\rm eff}(x_{\rm mb};\lambda,\beta)=1,
	\qquad
	\left.
	\frac{\partial V_{\rm eff}}{\partial x}
	\right|_{x_{\rm mb},\lambda,\beta}=0,
	\label{eq:mb_conditions}
\end{equation}
where the derivative is taken at fixed canonical angular momentum
$\lambda$ and fixed coupling $\beta$.

The energy condition fixes the positive mechanical angular momentum
as
$j_{\rm mb}=(2x_{\rm mb}+1)^2/
[\sqrt{2x_{\rm mb}}(2x_{\rm mb}-1)]$. Since
$j=\lambda+\beta/x$ on the present branch, the corresponding
canonical angular momentum is
$\lambda_{\rm mb}=j_{\rm mb}-\beta/x_{\rm mb}$.
Substituting these relations into the circular-orbit condition gives,
exactly within the adopted leading-dipole model,
\begin{equation}
	\sqrt{x_{\rm mb}}\,(2x_{\rm mb}+1)
	\left(4x_{\rm mb}^2-12x_{\rm mb}+1\right)
	=
	-2\sqrt{2}\,\beta\,(2x_{\rm mb}-1)^2 .
	\label{eq:mb_exact}
\end{equation}

The unsquared form is important because it retains the sign
information associated with the Lorentz-force orientation. Squaring
Eq.~\eqref{eq:mb_exact} yields
$x_{\rm mb}(2x_{\rm mb}+1)^2
(4x_{\rm mb}^2-12x_{\rm mb}+1)^2
=8\beta^2(2x_{\rm mb}-1)^4$, but also admits roots belonging to the
opposite sign of the original relation. Equation~\eqref{eq:mb_exact}
is therefore the appropriate expression for selecting the branch
considered here.

In the Schwarzschild limit, the relevant exterior solution is
\begin{equation}
	x_{\rm mb}^{(0)}
	=
	\frac{3+2\sqrt2}{2}
	\simeq2.91421 ,
	\label{eq:mb_schwarzschild}
\end{equation}
which corresponds, through
$r/M=x(1+1/2x)^2$, to the standard Schwarzschild marginally bound
radius $r_{\rm mb}=4M$.

For weak positive coupling, expansion about the Schwarzschild
solution gives
\begin{equation}
	x_{\rm mb}
	=
	x_{\rm mb}^{(0)}
	-\frac{\beta}{2}
	+\mathcal{O}(\beta^2),
	\qquad
	\beta\ll1 .
	\label{eq:mb_perturbative}
\end{equation}
The outward Lorentz interaction therefore shifts the marginally bound
orbit inward at first, in the same direction as the weak-coupling
radial stability boundary.

At larger coupling an important branch change occurs. The two
algebraic circular-orbit solutions $\lambda_\pm$ coalesce along the
marginally bound sequence at $x_{\rm mb}=3/2$ and
$\beta=2\sqrt3$. For $0\leq\beta<2\sqrt3$, the marginally bound
orbit lies on the $\lambda_+$ family continuously connected to the
positive-angular-momentum Schwarzschild sequence. Beyond the
coalescence point, the $\mathcal{E}=1$ continuation lies on the other
algebraic circular-orbit branch and should no longer be identified
with the outer Schwarzschild-connected $\lambda_+$ family.

The distinction is particularly important at strong coupling.
Equation~\eqref{eq:mb_exact} possesses a formal inner continuation
that approaches the isotropic horizon. Writing
$x_{\rm mb}=1/2+\delta$, with $\delta\ll1$, gives
\begin{equation}
	x_{\rm mb}
	=
	\frac{1}{2}
	+
	\frac{1}{\sqrt{2\beta}}
	+
	\mathcal{O}\!\left(\beta^{-1}\right),
	\qquad
	\beta\gg1 .
	\label{eq:mb_asymptotic}
\end{equation}
This is qualitatively different from the strong-coupling behaviour
of the inner fully stable orbit, which becomes vertically limited and
moves outward as $\beta^{2/3}$.

The near-horizon behaviour in
Eq.~\eqref{eq:mb_asymptotic} must, however, be interpreted only as a
formal property of the leading-dipole model. The electromagnetic
potential used throughout the charged-particle analysis is the
leading large-radius term of the relativistic exterior dipole field.
Its extrapolation to $x\simeq1/2$ therefore lies outside the regime
in which the electromagnetic approximation is quantitatively
controlled, even though the Schwarzschild gravitational sector
itself remains exact.

The marginally bound and fully stable inner orbits consequently probe
different parts of the charged-particle phase space. At weak coupling
both remain continuously connected to their Schwarzschild
counterparts and move inward. At stronger coupling their evolutions
separate: the fully stable boundary eventually becomes vertically
limited and moves outward, whereas the marginally bound solution
changes algebraic branch and formally approaches the near-horizon
region.

This contrast is itself sensitive to the Lorentz-force orientation.
As shown in Appendix~\ref{app:negative_beta}, on the complementary
inward-Lorentz branch the marginally bound orbit remains on the
Schwarzschild-connected positive-$j$ family and instead moves
outward at strong coupling, with the same $|\beta|^{2/3}$ radial
scaling characteristic of the radially marginal stability boundary
on that branch.

\subsection{Photon sphere of the Schwarzschild background}
\label{sec:photon_orbits}

Photon propagation is conceptually distinct from the Lorentz-force
dynamics of massive charged particles. The coupling
$\beta=e\mu/(mM^2)$ contains the specific charge of a massive
particle and has no analogue for a neutral photon. Moreover, within
the test-field approximation adopted here, the electromagnetic field
does not modify the Schwarzschild geometry. Vacuum photons therefore
follow the standard Schwarzschild null geodesics, independently of
$\beta$.

In Schwarzschild areal coordinates, the equatorial null radial
equation may be written as
$(dr/d\lambda_{\rm aff})^2+V_{\rm null}(r)=E_\gamma^2$, where
$\lambda_{\rm aff}$ is an affine parameter and
$V_{\rm null}(r)=(1-2M/r)L_\gamma^2/r^2$. Here $E_\gamma$ and
$L_\gamma$ are the conserved quantities associated with stationarity
and axial symmetry.

A circular null geodesic satisfies
$V_{\rm null}(r_{\rm ph})=E_\gamma^2$ together with
$dV_{\rm null}/dr=0$. These conditions give the standard unstable
Schwarzschild photon sphere at $r_{\rm ph}=3M$. The corresponding
critical impact parameter is
${\cal B}_{\rm ph}\equiv|L_\gamma|/E_\gamma=3\sqrt3\,M$.
Both quantities are determined entirely by the background geometry
and are therefore independent of the charged-particle coupling
$\beta$.

Since the charged-particle analysis is formulated in isotropic
coordinates, it is useful to express the same geometrical orbit in
terms of the isotropic radius. Using
$r=R(1+M/2R)^2$, the condition $r_{\rm ph}=3M$ gives two algebraic
roots. The one belonging to the exterior isotropic region $R>M/2$
is $R_{\rm ph}=M(1+\sqrt3/2)$, or equivalently
$x_{\rm ph}=1+\sqrt3/2\simeq1.86603$.

The second algebraic root,
$R/M=1-\sqrt3/2<1/2$, does not represent an additional physical
photon sphere in the Schwarzschild exterior. It arises from the
two-to-one character of the isotropic radial map when formally
extended beyond its exterior branch; indeed, the two isotropic
radii corresponding to the same areal radius are related by
$R\rightarrow M^2/(4R)$. The exterior value
$x_{\rm ph}=1+\sqrt3/2$ is therefore simply the usual Schwarzschild
photon sphere expressed in isotropic coordinates.

The external magnetic field can nevertheless affect observations
indirectly by modifying the dynamics and radiative properties of the
charged matter responsible for emission and absorption. Changes in
the particle distribution, velocity field, temperature, emissivity,
absorption, and polarization can alter the observed intensity and
polarization structure even though the underlying vacuum photon
trajectories remain unchanged. Plasma dispersion and polarized
radiative-transfer effects can additionally cause electromagnetic
wave propagation to depart from vacuum null-geodesic propagation.

Such matter-induced propagation effects should be distinguished from
a modification of the geometrical null-orbit structure itself. A
genuine shift of the vacuum light ring would require physics outside
the assumptions of the present model, such as electromagnetic
backreaction that changes the spacetime metric or a theory in which
photon characteristics do not coincide with the null cones of the
Schwarzschild geometry.

\subsection{Summary of analytical results}
\label{sec:analytical_summary}

For reference, Table~\ref{tab:analytical_summary} collects the
principal analytical results for the outward-Lorentz branch
$\beta>0$, with
$x=R/M$, $\lambda=L/(mM)$, and
$j=\lambda+\beta/x$. The Schwarzschild geometry is retained exactly,
while the electromagnetic interaction is described by the leading
asymptotic dipole potential.

\newpage 
\begin{widetext}
\begin{table}[htbp!]
	\centering
	\setlength{\arraycolsep}{6pt} 
	\renewcommand{\arraystretch}{1.6} 
	
	\begin{minipage}{\textwidth}
		\centering
		\caption{Principal analytical results for the
			outward-Lorentz charged-particle branch $\beta>0$ in the
			isotropic-coordinate leading-dipole model.}
		\label{tab:analytical_summary}
		
		\vspace{10pt}
		
		\centerline{$\begin{array}{|c|c|}
	\hline
			\textbf{Quantity} & \textbf{Result} \\
			\hline
			\text{Mechanical angular momentum} & \displaystyle j=\lambda+\frac{\beta}{x} \\
			\hline
			\text{Effective potential} & \displaystyle V_{\rm eff} = \Phi^2 \left[ 1+\frac{\left(\lambda+\frac{\beta}{x}\right)^2}{\Psi^4x^2} \right] \\
			\hline
			\text{Schwarzschild-connected circular branch} & \lambda=\lambda_+(x,\beta), \text{ via Eq.}~\eqref{eq:lambda_circ} \\
			\hline
			\text{Radial stability} & \Omega_r^2>0; \text{ determined by Eq.}~\eqref{eq:isco_polynomial} \\
			\hline
			\text{Vertical stability} & \displaystyle \frac{\Omega_\theta^2}{\Omega_\phi^2} = 1-\frac{2\beta}{jx}, \quad j>\frac{2\beta}{x} \\
			\hline
			\begin{array}{l} \text{Inner boundary of} \\ \text{fully stable motion} \end{array} & 
			\begin{array}{l} \text{Radially limited for } \beta<\beta_{\rm c}\text{, vertically for } \beta>\beta_{\rm c}; \\ \beta_{\rm c}\simeq3.05318 \end{array} \\
			\hline
			\begin{array}{l} \text{Weak-coupling radial} \\ \text{stability boundary} \end{array} & \displaystyle x_{\rm rISCO} = \frac{5+2\sqrt6}{2} - \left( \frac{3\sqrt3}{8} +\frac{\sqrt2}{4} \right)\beta +\mathcal{O}(\beta^2) \\
			\hline
			\begin{array}{l} \text{Strong-coupling inner} \\ \text{stability boundary} \end{array} & \displaystyle x_{\rm ISCO} \sim 6^{1/3}\beta^{2/3} \text{ (vertical stability)} \\
			\hline
			\text{Marginally bound orbit} & \text{Eq.}~\eqref{eq:mb_exact}; \displaystyle x_{\rm mb} = \frac{3+2\sqrt2}{2} -\frac{\beta}{2} +\mathcal{O}(\beta^2) \ \ (\beta\ll1) \\
			\hline
			\begin{array}{l} \text{Formal strong-coupling} \\ \text{marginally bound continuation} \end{array} & \displaystyle x_{\rm mb} = \frac12 +\frac{1}{\sqrt{2\beta}} +\mathcal{O}(\beta^{-1}); \text{ inner branch} \\
			\hline
			\text{Schwarzschild photon sphere} & \displaystyle r_{\rm ph}=3M, \quad x_{\rm ph}=1+\frac{\sqrt3}{2} \simeq1.86603 \\
			\hline
			\text{Critical photon impact parameter} & \displaystyle {\cal B}_{\rm ph}=3\sqrt3\,M \text{ (independent of } \beta) \\
			\hline
		\end{array}$}
	\end{minipage}
\end{table}
\end{widetext}

In the Schwarzschild limit the characteristic radii obey
\begin{equation}
	x_{\rm ph}
	<
	x_{\rm mb}^{(0)}
	<
	x_{\rm ISCO}^{(0)},
\end{equation}
corresponding to the familiar areal radii $3M$, $4M$, and $6M$.
At finite magnetic coupling, however, the charged-particle orbital
structure no longer follows this simple Schwarzschild pattern.

On the $\beta>0$ branch, weak outward Lorentz coupling shifts both
the marginally bound orbit and the radial stability boundary inward.
At stronger coupling, vertical stability becomes the limiting
condition and the inner boundary of fully stable circular motion
moves outward. The marginally bound continuation instead changes
algebraic branch and, formally within the leading-dipole model,
evolves toward the near-horizon region.

The same distinction appears in the epicyclic dynamics. Vertically
stable orbits satisfy $\Omega_\theta<\Omega_\phi$, while the vertical
mode vanishes at a vertically marginal boundary. These
charged-particle effects remain separate from the null-geodesic
structure of the fixed Schwarzschild spacetime, for which neither the
photon-sphere radius nor the vacuum critical impact parameter depends
on $\beta$.

\section{Astrophysical applications}
\label{sec:applications}

The analytical results derived above provide a framework for
assessing the motion of charged test particles in magnetized
compact-object environments. Their astrophysical interpretation,
however, requires several qualifications. Most importantly, the
dimensionless magnetic coupling depends not only on the external
field but also on the specific charge of the orbiting particle or
structure. Electrons, protons, ions, charged grains, and weakly
charged macroscopic bodies can therefore probe very different
dynamical regimes in the same magnetosphere.

The present model should accordingly be interpreted as a
test-particle description of charged tracers moving in an externally
specified dipolar field. It is not a model of the bulk dynamics of a
quasi-neutral accretion flow, for which pressure, currents,
collective electromagnetic effects, magnetic stresses, radiation,
collisions, and relativistic magnetohydrodynamics may all be
important.

For a neutron star, the dipolar field may be associated with the
stellar magnetic moment, while the stellar surface provides an
additional inner boundary for particle motion. For a black hole, by
contrast, a stationary dipole-like field must be supported by
external sources, such as currents in the surrounding plasma or
accretion flow; it should not be interpreted as an intrinsic
magnetic dipole moment of a Schwarzschild black hole. In either case,
the test-field approximation requires the electromagnetic
stress-energy to remain sufficiently small for the Schwarzschild
geometry to provide an adequate fixed background.

In geometric units, the signed coupling is
$\beta=e\mu/(mM^2)$. We continue to focus on positive azimuthal
motion on the outward-Lorentz branch $\beta>0$; the complementary
orientation is analysed in Appendix~\ref{app:negative_beta}. When
only the coupling strength is relevant, we use $|\beta|$. This
particle-dependent parameter affects massive charged-particle
dynamics but not the vacuum null-geodesic structure of the fixed
Schwarzschild background.

\subsection{Parameter space of astrophysical systems}
\label{sec:parameter_space}

In geometric units, the coupling magnitude is
$|\beta|=|e\mu|/(mM^2)$, where
$M=GM_\star/c^2$ is the gravitational length of the central object.
Thus $|\beta|$ depends on the test body's specific charge, the
magnetic dipole moment, and the mass scale of the gravitating source.

With the dipole normalization adopted throughout this work, the
corresponding Gaussian-cgs expression is
\begin{equation}
	|\beta|
	=
	\frac{|q|\,\mu_{\rm cgs}}
	{m c^2 M^2},
	\qquad
	M=\frac{GM_\star}{c^2},
	\label{eq:b_cgs}
\end{equation}
where $q$ and $m$ are the physical charge and rest mass of the test
body and $\mu_{\rm cgs}$ is the magnitude of the magnetic dipole
moment. The sign of $\beta$ is fixed by the relative orientation of
the charge, dipole moment, and chosen azimuthal direction.

For order-of-magnitude estimates of a dipolar stellar field one may
write
$\mu_{\rm cgs}\simeq B_{\rm eq}r_\star^3
\simeq B_{\rm p}r_\star^3/2$, where $r_\star$ is the stellar areal
radius and $B_{\rm eq}$ and $B_{\rm p}$ are characteristic
equatorial and polar surface-field strengths. These expressions are
leading dipole estimates rather than exact relativistic relations
between locally measured surface fields and the asymptotic magnetic
moment.

Equation~\eqref{eq:b_cgs} makes clear that no unique value of
$|\beta|$ can be assigned to a magnetized compact object. At fixed
external field, $|\beta|\propto|q|/m$, so different charged species
may experience radically different Lorentz-force dynamics.

As an illustration, for an electron near a neutron star with
$M_\star=1.4\,M_\odot$, areal radius $r_\star=12\,{\rm km}$, and
equatorial surface field $B_{\rm eq}=10^{12}\,{\rm G}$,
Eq.~\eqref{eq:b_cgs} gives
\begin{equation}
	|\beta_e|
	\simeq
	2.4\times10^{16}
	\left(
	\frac{B_{\rm eq}}{10^{12}\,{\rm G}}
	\right)
	\left(
	\frac{r_\star}{12\,{\rm km}}
	\right)^3
	\left(
	\frac{1.4\,M_\odot}{M_\star}
	\right)^2 .
	\label{eq:b_electron_NS}
\end{equation}
For a proton in the same field,
$|\beta_p|\simeq1.3\times10^{13}$, while increasing the magnetic
field from $10^{12}\,{\rm G}$ to $10^{15}\,{\rm G}$ increases both
estimates by three orders of magnitude.

These very large values arise primarily from the enormous specific
charges of elementary particles. They should not be interpreted as
evidence that electromagnetic backreaction on the spacetime is
large. The test-field approximation is controlled by the
stress-energy of the electromagnetic field relative to that of the
gravitating source, whereas $\beta$ additionally contains the
particle-dependent factor $q/m$. A particle may therefore be in an
extremely strong Lorentz-force regime while the electromagnetic field
still produces a negligible modification of the background metric.

Large elementary-particle values of $|\beta|$ do, however, expose
other limitations of the conservative single-particle description.
On the outward-Lorentz branch, the inner boundary of fully stable
circular motion scales as $x_{\rm ISCO}\propto\beta^{2/3}$ at strong
coupling and can therefore move to large radii. Over such distances,
the actual magnetosphere need not remain globally dipolar. Radiation
reaction, electric fields, collisions, charge screening, and
collective plasma effects may also compete with or dominate the
conservative Lorentz-force dynamics of elementary particles.

Conversely, weakly charged macroscopic bodies or suitably idealized
coherent structures can possess much smaller effective specific
charges and may correspond to $|\beta|$ of order unity or moderately
larger even in a strong field. Such cases provide a more direct
setting in which the intermediate-coupling circular-orbit and
epicyclic structure derived above may be useful, although a coherent
plasma element should not in general be identified literally with a
single point charge.

For neutron stars, the material surface imposes an additional
constraint. Stellar radii are normally quoted as areal radii,
whereas the orbital analysis uses the isotropic coordinate $R$. The
corresponding isotropic surface radius is
$R_{\star,\rm iso}
=[r_\star-M+\sqrt{r_\star(r_\star-2M)}]/2$.
Only orbits with $R>R_{\star,\rm iso}$ are physically exterior to
the star.

For the fiducial values $M_\star=1.4\,M_\odot$ and
$r_\star=12\,{\rm km}$, one obtains
$M\simeq2.067\,{\rm km}$,
$R_{\star,\rm iso}\simeq9.824\,{\rm km}$, and
$x_\star=R_{\star,\rm iso}/M\simeq4.752$. The relevant physical inner
radius is therefore determined by the larger of the stellar surface
and the dynamical stability boundary. A formal charged-particle ISCO
with $x_{\rm ISCO}<x_\star$ lies inside the star and cannot delimit
an exterior circular-orbit sequence.

This neutron-star interpretation also inherits the assumptions of
the background geometry. A Schwarzschild exterior corresponds to an
idealized non-rotating, spherically symmetric object. Rapid rotation,
frame dragging, stellar multipole moments, and departures of the
magnetic field from a stationary aligned dipole require a more
general treatment.

For black holes there is no material surface, but the external
magnetic field must be supported by sources outside the hole. Its
physical domain is therefore determined by the current distribution
and magnetospheric structure. In addition, the electromagnetic
potential used in the charged-particle analysis is the leading
large-radius term of the relativistic dipole field and is
quantitatively controlled only in its asymptotic regime. Predictions
obtained by extrapolating this approximation into the immediate
near-horizon region should consequently be regarded with caution.
By contrast, the strong-$\beta$ fully stable boundary moves toward
$x\gg1$, where the local use of the leading asymptotic dipole term
becomes progressively better controlled, even though an actual
astrophysical magnetosphere need not remain globally dipolar over
such large scales.

Equation~\eqref{eq:b_cgs} applies equally to supermassive black-hole
environments once an external magnetic configuration and the
specific charge of the test object have been specified. There is,
however, no universal scaling of $\beta$ with black-hole mass
without an additional prescription for how the characteristic
magnetic moment varies between systems. Only if $\mu$ were held
fixed would one obtain $|\beta|\propto M^{-2}$.

More generally, assigning the elementary-particle value of $|q|/m$
to the bulk motion of a quasi-neutral plasma is inappropriate. The
orbital frequencies and stability boundaries derived here are most
naturally interpreted as properties of charged tracers or suitably
idealized weakly charged structures, rather than as direct
predictions for an entire accretion flow.

These considerations motivate treating $\beta$ primarily as a
dimensionless control parameter. A concrete astrophysical
application additionally requires specification of the central mass,
the strength, geometry, and radial extent of the external
electromagnetic field, the location of any material surface, and,
crucially, the effective specific charge and Lorentz-force
orientation of the orbiting particle or structure.

\subsection{Magnetars}
\label{sec:magnetars}

Magnetars provide a natural setting in which electromagnetic forces
can strongly influence charged-particle motion. Their inferred
large-scale magnetic fields are commonly associated with the
$10^{14}$--$10^{15}\,{\rm G}$ range, although the phenomenological
magnetar class also includes objects with substantially weaker
inferred external dipole fields. The dynamical coupling $\beta$,
however, is not determined by the magnetar alone: as emphasized in
Sec.~\ref{sec:parameter_space}, it depends linearly on the effective
specific charge $q/m$ of the orbiting particle or structure.
Throughout this subsection we continue to consider positive
azimuthal motion on the outward-Lorentz branch $\beta>0$.

For reference, consider an idealized non-rotating neutron star with
$M_\star\simeq1.4\,M_\odot$ and areal radius
$r_\star\simeq12\,{\rm km}$. Its gravitational length is
$M=GM_\star/c^2\simeq2.07\,{\rm km}$. Using the isotropic--areal
radius relation, the stellar surface lies at
$x_\star\simeq4.75$. The charged-particle circular-orbit analysis is
therefore physically applicable only outside this radius.

The location of the dynamical stability boundary on the $\beta>0$
branch is non-monotonic. At weak coupling, radial marginal stability
determines the formal inner boundary of fully stable circular motion,
and the outward Lorentz force shifts it inward relative to the
Schwarzschild value. For example, at $\beta=1$ the formal stability
boundary lies inside the surface of the representative star, so the
stellar surface rather than the charged-particle ISCO provides the
physical inner cutoff for exterior circular motion.

At stronger coupling, vertical stability becomes the limiting
condition and the fully stable boundary moves outward again. For the
representative couplings $\beta=10$ and $100$ considered in
Table~\ref{tab:circular_catalogue}, the corresponding stability
boundaries lie outside the fiducial stellar surface. Within the
present idealized model, the relevant exterior inner radius is thus
the larger of $x_\star$ and $x_{\rm ISCO}$, where
$x_{\rm ISCO}$ denotes the innermost member of the fully stable
circular-orbit sequence, whether determined by radial or vertical
marginality.

In the strong-coupling regime, vertical marginal stability gives
\begin{equation}
	x_{\rm ISCO}
	\sim
	6^{1/3}\beta^{2/3},
	\qquad
	\beta\gg1 .
	\label{eq:magnetar_isco_scaling}
\end{equation}
The displacement toward $x\gg1$ improves the local validity of the
leading large-radius dipole approximation used in the analytical
model. Its astrophysical interpretation nevertheless requires
caution: for the very large specific charges of elementary
particles, the formal stability boundary may occur at radii over
which a stationary, globally dipolar magnetosphere is no longer a
realistic description.

Real magnetar magnetospheres contain plasma and currents and need not
be either magnetostatic or purely dipolar. Rotation can generate
electric fields, while screening, collisions, radiation reaction,
and collective electromagnetic effects may compete with the
conservative single-particle dynamics. The scaling
\eqref{eq:magnetar_isco_scaling} should therefore be interpreted as a
property of the idealized Schwarzschild--dipole model rather than as
a direct prediction for the inner structure of a magnetar
magnetosphere.

For any specified effective coupling, the dimensionless frequencies
derived above can be converted to physical cyclic frequencies through
$\nu_i=c^3\bar\Omega_i/(2\pi GM_\star)$, with
$i=\phi,r,\theta$ and $\bar\Omega_i=M\Omega_i$. As a reference, the
neutral Schwarzschild ISCO has
$\bar\Omega_\theta=\bar\Omega_\phi=6^{-3/2}\simeq0.0680$, which for
$M_\star=1.4\,M_\odot$ corresponds to a cyclic frequency of order
$1.6\,{\rm kHz}$.

On the outward-Lorentz branch, however, the frequencies associated
with the formal inner stability boundary evolve non-monotonically
with $\beta$. At weak coupling the radially marginal orbit moves
inward and its orbital frequency increases, although for a neutron
star this orbit may lie below the stellar surface. Once vertical
marginality becomes the limiting condition, the inner stable radius
moves outward and its orbital frequency decreases.

At a vertically marginal inner boundary,
$\Omega_\theta(x_{\rm ISCO})=0$ exactly. This vanishing of the
linear vertical epicyclic frequency identifies a soft mode and should
not be interpreted as an observable zero-frequency oscillation. In
the strong-coupling limit the orbital frequency instead behaves as
$M\Omega_{\phi,\rm ISCO}\sim(3\beta)^{-1}$.

The single-particle frequencies derived here should not be identified
directly with quasi-periodic oscillations observed in magnetar
emission. Magnetar QPOs are generally associated with collective
stellar and magnetospheric dynamics, including crustal elasticity,
Alfv\'en motion, and magneto-elastic coupling. The present
frequencies characterize instead the local orbital dynamics of
charged test particles or suitably idealized weakly charged
structures. Relating such frequencies to an observed modulation would
require a specific excitation and emission mechanism together with
radiative transfer.

Likewise, the vertical epicyclic frequency should not be interpreted
as determining a hydrostatic magnetospheric scale height. For an
individual particle with specified vertical oscillator energy, the
proper meridional excursion is given by
Eq.~\eqref{eq:vertical_excursion}. As vertical marginal stability is
approached, the harmonic estimate grows as
$\omega_{\theta,s}^{-1}$ and eventually breaks down. The vertical
structure of a collective plasma is additionally controlled by
pressure, currents, magnetic stresses, radiation, and the global
magnetospheric configuration.

Dissipative effects may also become important for elementary charged
particles in magnetar-strength fields. The conservative circular
solutions obtained here are therefore most naturally regarded as
reference trajectories about which dissipative or collective
evolution can occur, rather than as indefinitely stationary particle
configurations. This interpretation provides a natural connection
with the radiation-reaction analysis developed in the companion
work~\cite{CRSP_react}.

\subsection{Stellar-mass black holes in strong external magnetic fields}
\label{sec:black_holes}

The charged-particle dynamics derived above can also be applied
formally to a stellar-mass black hole immersed in an externally
supported magnetic field. In the Schwarzschild model considered
here, the black hole possesses no intrinsic magnetic dipole; the
field must instead be generated by external sources, such as currents
in the surrounding plasma or accretion flow. It is treated as a test
field whose stress-energy is assumed not to modify appreciably the
background geometry.

As in the neutron-star case, the dimensionless coupling
$\beta=e\mu/(mM^2)$ is particle dependent and cannot be assigned
uniquely to the black-hole environment itself. The same external
field can correspond to extremely large $|\beta|$ for elementary
particles and much smaller effective couplings for weakly charged
macroscopic bodies or suitably idealized coherent structures.
Throughout this subsection we consider positive azimuthal motion on
the outward-Lorentz branch $\beta>0$.

The inner boundary of fully stable circular motion depends
non-monotonically on the coupling. For
$\beta<\beta_{\rm c}$, radial marginal stability sets the boundary
and the outward Lorentz interaction shifts it inward relative to the
Schwarzschild value. At
$\beta=\beta_{\rm c}\simeq3.05318$, the radial and vertical
marginal-stability curves intersect and both epicyclic modes become
marginal. For $\beta>\beta_{\rm c}$, vertical stability instead
determines the inner boundary, which then moves outward.

In the strong-coupling regime, the vertically marginal boundary
obeys
\begin{equation}
	x_{\rm ISCO}
	\sim
	6^{1/3}\beta^{2/3},
	\qquad
	\beta\gg1 .
	\label{eq:risco_bh_scaling}
\end{equation}
Since $x_{\rm ISCO}\gg1$ in this limit, the areal radius satisfies
$r_{\rm ISCO}/M=x_{\rm ISCO}+1+\mathcal{O}(x_{\rm ISCO}^{-1})$,
so that $r_{\rm ISCO}\simeq Mx_{\rm ISCO}$ at leading order.

Vertical marginal stability also gives
$j_{\rm ISCO}=2\beta/x_{\rm ISCO}$. Combining this relation with
Eq.~\eqref{eq:risco_bh_scaling} and the circular-orbit frequency
yields
\begin{equation}
	M\Omega_{\phi,\rm ISCO}
	\sim
	\frac{1}{3\beta},
	\qquad
	\beta\gg1 .
	\label{eq:Omega_bh_strong}
\end{equation}
The corresponding orbital period measured with respect to the
asymptotic Schwarzschild time is therefore
\begin{equation}
	P_{\rm ISCO}
	\sim
	6\pi\beta\,
	\frac{GM_\star}{c^3}.
	\label{eq:period_bh}
\end{equation}
These scalings follow from the magnetically modified circular-orbit
conditions themselves; an independent Keplerian prescription should
not be imposed on the charged-particle sequence.

For very large elementary-particle couplings, the formal stability
boundary~\eqref{eq:risco_bh_scaling} may occur far from the black
hole. This has two complementary implications. Locally, the
large-radius displacement improves the validity of the leading
asymptotic dipole term used in the analytical model. Globally,
however, there is no reason for an astrophysical black-hole
magnetosphere to remain stationary, vacuum, and purely dipolar over
an arbitrarily large radial range. The current distribution, induced
electric fields, plasma loading, reconnection, radiation reaction,
and interaction with the accretion flow may then become more
important than the idealized conservative test-particle dynamics.

A large value of $|\beta|$ should also be distinguished from strong
electromagnetic backreaction on the spacetime. The former contains
the particle-dependent factor $|e|/m$ and may be enormous even when
the electromagnetic stress-energy remains sufficiently small for the
Schwarzschild test-field approximation to remain accurate.

For a specified effective coupling, the physical orbital and
epicyclic frequencies are obtained from
\begin{equation}
	\nu_i
	=
	\frac{c^3}{2\pi G M_\star}\,
	\bar\Omega_i,
	\qquad
	i=\phi,r,\theta ,
	\label{eq:bh_frequency_conversion}
\end{equation}
where $\bar\Omega_i=M\Omega_i$ denotes the corresponding
dimensionless coordinate-time frequency evaluated on the
charged-particle circular-orbit sequence.

The nature of the inner stability boundary is reflected directly in
the epicyclic spectrum. For $\beta<\beta_{\rm c}$, the limiting
orbit is radially marginal, so $\Omega_r=0$ while
$\Omega_\theta>0$. At $\beta=\beta_{\rm c}$ both modes vanish
simultaneously. For $\beta>\beta_{\rm c}$, the inner boundary is
vertically marginal, with $\Omega_\theta=0$ while $\Omega_r>0$.
Strong outward coupling therefore changes the identity of the soft
mode from radial to vertical.

These characteristic frequencies describe the dynamics of charged
test particles and should not be identified directly with observed
quasi-periodic variability. Any such connection would require a
mechanism for exciting the relevant modes, nonlinear coupling,
dissipation, emission, and radiative transfer.

More fundamentally, the charged-particle stability boundary obtained
here should not be identified with the inner edge of a realistic
black-hole accretion flow. The calculation follows a single test body
with a prescribed charge-to-mass ratio, whereas an astrophysical
accretion flow is approximately quasi-neutral and evolves
collectively under pressure, currents, magnetic stresses,
turbulence, dissipation, and relativistic magnetohydrodynamics.
Substituting the electron or proton value of $e/m$ into $\beta$ does
not therefore determine an accretion-disk inner radius.

For the same reason, the present calculation does not by itself
predict a modified disk radiative efficiency, nor can an
observationally inferred inner disk radius be translated directly
into a constraint on $\beta$ or on the external magnetic field.
Such connections require a self-consistent plasma model together
with an appropriate emission and radiative-transfer calculation.

The results are instead most naturally interpreted as describing
charged tracers, weakly charged bodies, or suitably idealized
coherent structures whose motion can be treated independently of the
bulk plasma. Within this domain, the principal effect of the outward
Lorentz interaction is the qualitative change in stability structure:
weak coupling shifts the radially marginal boundary inward, whereas
sufficiently strong coupling produces a vertically marginal inner
boundary that moves outward and is associated with progressively
lower orbital frequencies.

Finally, the Schwarzschild background itself is an idealization for
astrophysical black holes. Rotation introduces frame dragging and
modifies both the orbital structure and the admissible
electromagnetic configurations. Extending the present analysis to a
rotating background is therefore necessary before applying these
results quantitatively to realistic stellar-mass black-hole systems.

\subsection{Supermassive black holes}
\label{sec:supermassive}

The charged-particle framework developed above can also be applied
formally to supermassive black holes immersed in externally supported
magnetic fields. As in the stellar-mass case, its interpretation
requires care because the dimensionless coupling
$\beta=e\mu/(mM^2)$ depends both on the electromagnetic
configuration and on the specific charge of the orbiting particle or
structure. Throughout this subsection we continue to consider
positive azimuthal motion on the outward-Lorentz branch $\beta>0$.

For a black hole with $M_\star\sim10^8\,M_\odot$, the gravitational
length $M=GM_\star/c^2$ is of order
$1.5\times10^{13}\,{\rm cm}$. If the physical dipole moment $\mu$
and the particle specific charge are held fixed, then
$|\beta|\propto(|q|/m)|\mu|M^{-2}$ and the coupling decreases as
$M_\star^{-2}$. This is not, however, a universal mass scaling for
astrophysical black holes, because the magnetic moment itself depends
on the strength, size, and current distribution of the external
magnetosphere.

This dependence is particularly transparent if the field is specified
at a reference radius $R_0$. Within the leading dipole approximation,
$\mu_{\rm cgs}\simeq B_0R_0^3$. Writing
$R_0=\kappa M$ and using Eq.~\eqref{eq:b_cgs} then gives
$|\beta|\simeq |q|B_0\kappa^3M/(mc^2)$. Thus, if the characteristic
field strength is compared at a fixed dimensionless radius
$\kappa=R_0/M$, one instead obtains
$|\beta|\propto B_0M$ for fixed specific charge. More generally,
there is no universal scaling of $\beta$ with black-hole mass
without specifying how both the magnetic field strength and its
characteristic spatial scale vary between systems.

This also emphasizes that $\beta$ is not a property of the black hole
or magnetosphere alone. Elementary particles, ions, weakly charged
macroscopic bodies, and nearly neutral coherent structures can
correspond to very different values of $\beta$ in the same external
field.

For sufficiently strong outward coupling, the inner boundary of
fully stable circular motion is vertically marginal and obeys
\begin{equation}
	x_{\rm ISCO}
	\sim
	6^{1/3}\beta^{2/3},
	\qquad
	\beta\gg1 .
	\label{eq:smbh_isco_scaling}
\end{equation}
Large particle couplings can therefore displace the formal fully
stable inner orbit to many gravitational radii from the black hole.

Such large-radius solutions require a careful physical
interpretation. Their displacement toward $x\gg1$ improves the local
validity of the leading asymptotic dipole expansion used in the
analytical calculation. At the same time, an actual black-hole
magnetosphere need not remain stationary, vacuum, and purely dipolar
over the corresponding radial range. Large-scale currents, the
accretion flow, jets, reconnection, turbulence, electric fields, and
collective plasma effects can substantially modify the field
configuration, while dissipative effects may become important for
elementary particles. Equation~\eqref{eq:smbh_isco_scaling} should
therefore be interpreted as an asymptotic property of the idealized
charged-particle model rather than as a prediction for the inner edge
of an astrophysical accretion flow.

A large value of $|\beta|$ likewise does not imply strong
electromagnetic backreaction on the spacetime. The particle coupling
contains the specific charge $|q|/m$, whereas the validity of the
test-field approximation is governed by the stress-energy of the
electromagnetic field. An elementary particle may therefore occupy a
very strongly coupled Lorentz-force regime while the Schwarzschild
metric remains an excellent approximation to the background
geometry.

For a specified dimensionless coupling and circular orbit, the
physical orbital and epicyclic frequencies follow from
\begin{equation}
	\nu_i
	=
	\frac{c^3}{2\pi G M_\star}\,
	\bar\Omega_i,
	\qquad
	i=\phi,r,\theta .
	\label{eq:smbh_frequency_conversion}
\end{equation}
Thus, for fixed dimensionless orbital parameters, the dimensional
frequency scale varies as $M_\star^{-1}$. Across different
astrophysical systems, however, the dimensionless frequencies need
not be mass independent because $\beta$ itself can vary with the
magnetic configuration and central mass.

The nature of the inner stability boundary is reflected directly in
the epicyclic spectrum. For $\beta<\beta_{\rm c}$ the limiting orbit
is radially marginal, with $\Omega_r=0$ and
$\Omega_\theta>0$. At $\beta=\beta_{\rm c}$ the radial and vertical
marginal curves intersect and both frequencies vanish. For
$\beta>\beta_{\rm c}$, including the strong-coupling regime of
Eq.~\eqref{eq:smbh_isco_scaling}, the inner boundary is vertically
marginal, so $\Omega_\theta=0$ while $\Omega_r>0$. The
strong-coupling inner orbit is therefore characterized by a soft
vertical rather than radial mode.

Magnetic fields may nevertheless play an important role in
horizon-scale observations, but this should be distinguished from a
shift of the vacuum photon sphere. As discussed in
Sec.~\ref{sec:photon_orbits}, the Schwarzschild photon sphere remains
at $r_{\rm ph}=3M$, with critical vacuum impact parameter
${\cal B}_{\rm ph}=3\sqrt3\,M$, independently of the charged-particle
coupling in the present test-field approximation.

The external field can instead influence an observed image by
altering the physical state of the emitting and absorbing plasma.
Particle densities and velocities, synchrotron emission and
absorption, polarization, and Faraday effects all depend on the
plasma and magnetic configuration. Substantial changes in brightness
and polarization structure can therefore occur without any change in
the underlying vacuum photon sphere.

If dispersive plasma effects become important, electromagnetic-wave
propagation itself may depart from vacuum null-geodesic propagation.
Such effects belong to the optical properties of the intervening
medium and should be distinguished from a modification of the
geometrical light ring of the spacetime.

A genuine displacement of the vacuum null-orbit structure would
require physics outside the assumptions of the present model, such
as electromagnetic backreaction that modifies the metric or a theory
in which photon characteristics do not coincide with the null cones
of the Schwarzschild geometry.

The main relevance of the present calculation to supermassive
black-hole environments is therefore the separation between
charged-matter dynamics and vacuum photon propagation. An external
magnetic field can strongly reorganize charged-particle motion and
the radiative properties of the surrounding plasma while leaving the
vacuum Schwarzschild null-geodesic structure unchanged. Quantitative
applications to horizon-scale observations require a
self-consistent magnetospheric or relativistic plasma model together
with polarized radiative transfer and, for realistic astrophysical
black holes, the effects of rotation.

\subsection{Summary of astrophysical implications}
\label{sec:applications_summary}

The principal astrophysical implications of the
Schwarzschild--dipole model are summarized in
Table~\ref{tab:applications_summary}. They should be interpreted
within the test-particle, test-field, and leading-dipole
approximations adopted throughout this work.

\begin{table*}[th!]
	\centering
	\caption{Summary of the principal astrophysical implications of the Schwarzschild--dipole charged-particle model. The outward-Lorentz branch $\beta>0$ is assumed.}
	\label{tab:applications_summary}
	
	\renewcommand{\arraystretch}{1.35}
	\setlength{\extrarowheight}{3pt}
	\setlength{\tabcolsep}{1pt}
	
	\begin{ruledtabular}
		\begin{tabular}{|c|c|c|}
			
			\textbf{System or effect}
			&
			\textbf{Main implication}
			&
			\textbf{Principal qualification}
			\\
			
			\hline
			
			Magnetars
			&
			\parbox[t]{6.5cm}{The Lorentz force can strongly modify charged-particle orbits. Weak outward coupling shifts the radial stability boundary inward, whereas sufficiently strong coupling makes vertical marginal stability the limiting condition and gives $x_{\rm ISCO}\sim6^{1/3}\beta^{2/3}$ within the adopted model.}
			&
			\parbox[t]{7.0cm}{The stellar surface may lie outside the formal dynamical stability boundary. For large elementary-particle couplings, radiation reaction, plasma effects, electric fields, and the actual magnetospheric geometry may dominate over the idealized conservative dynamics.}
			\\
\hline
			
			Stellar-mass black holes
			&
			\parbox[t]{6.5cm}{An externally supported magnetic field can substantially modify the equilibrium, stability, and characteristic frequencies of charged tracers.}
			&
			\parbox[t]{7.0cm}{The charged-particle stability boundary is not, in general, the inner edge of a quasi-neutral accretion flow. A realistic disk requires a collective plasma or GRMHD description.}
			\\
	\hline		
			Supermassive black holes
			&
			\parbox[t]{6.5cm}{Magnetic fields can strongly affect the dynamics and radiative properties of charged matter while leaving the vacuum Schwarzschild photon sphere unchanged in the test-field approximation.}
			&
			\parbox[t]{7.0cm}{Horizon-scale observables depend on the plasma dynamics, emissivity, absorption, polarization, Faraday effects, dispersion, and radiative transfer rather than on a $\beta$-dependent displacement of the geometrical photon sphere.}
			\\
	\hline		
			Epicyclic commensurabilities
			&
			\parbox[t]{6.5cm}{The radial and vertical frequencies cross low-order rational ratios at characteristic radii. Radially limited sequences contain sub-unity crossings, whereas vertically limited sequences begin with $\Omega_r/\Omega_\theta\rightarrow\infty$, cross super-unity ratios, and may subsequently cross unity.}
			&
			\parbox[t]{7.0cm}{A frequency commensurability identifies only a candidate location for nonlinear mode coupling. Resonance widths, capture, secular energy transfer, and observable variability require a nonlinear and generally dissipative analysis.}
			\\
			
		\end{tabular}
	\end{ruledtabular}
	
	\renewcommand{\arraystretch}{1}
	\setlength{\extrarowheight}{0pt}
\end{table*}

A central point in all of these applications is that, in geometric
units, $\beta=e\mu/(mM^2)$, is a signed, particle-dependent coupling rather than a property of
the compact object alone. Its magnitude depends on the effective
specific charge of the orbiting component and on the external
magnetic moment, while its sign also depends on the relative
orientation of the charge, field, and azimuthal motion. A numerical
value of $\beta$ therefore acquires astrophysical meaning only once
the nature of the charged particle or structure and the magnetic
configuration have been specified.

A large value of $|\beta|$ should likewise not be confused with
strong electromagnetic backreaction on the spacetime. The former
contains the specific charge $|e|/m$, whereas the latter is governed
by the stress-energy of the electromagnetic field. Very strong
charged-particle Lorentz coupling can therefore coexist with a
well-controlled test-field geometry.

Within these qualifications, the Schwarzschild--dipole system
provides a useful analytical laboratory for separating three
conceptually distinct sectors: the fixed gravitational geometry, the
Lorentz-force dynamics of massive charged particles, and the
collective plasma and radiative processes through which those
dynamics might become observable. The external field can reorganize
the charged-particle circular-orbit and stability structure without
altering the vacuum Schwarzschild photon sphere.

\section{Conclusions}
\label{sec:conclusions}

We have developed an analytical description of the conservative
motion of relativistic charged test particles in a Schwarzschild
spacetime endowed with an externally supported dipolar magnetic
field. The Schwarzschild geometry has been retained exactly in
isotropic coordinates, while the electromagnetic sector has been
described by the leading asymptotic term of the relativistic exterior
dipole potential. Within this framework, we have obtained the
effective potential for equatorial motion, the circular-orbit family,
the radial and vertical epicyclic frequencies, and the conditions
governing marginal stability and marginal binding.

A central point of the analysis is the distinction between canonical
and mechanical angular momentum. With the conventions adopted in this
work, the signed coupling
$\beta=e\mu/(mM^2)$ enters the mechanical angular momentum through
$j=\lambda+\beta/x$. For positive azimuthal motion, the branch
$\beta>0$ corresponds to an outward Lorentz force. This sign
convention is important, since reversing the electromagnetic
orientation changes not only the force balance but also the stability
properties of the circular-orbit sequence.

The most significant dynamical result is that the inner boundary of
stable circular motion is controlled by two competing instability
mechanisms. At weak positive coupling, radial marginal stability
continues to determine the limiting circular orbit, and the outward
Lorentz interaction shifts this boundary inward relative to the
Schwarzschild case. As the coupling increases, however, the vertical
restoring force is progressively weakened. Beyond a transition
between the radial and vertical stability boundaries, vertical
marginal stability becomes the relevant limiting condition and the
inner stable orbit moves outward. The strong-coupling regime is
therefore not simply a stronger version of the weak-coupling radial
ISCO displacement, but represents a qualitatively different
stability regime.

This transition is reflected directly in the epicyclic frequencies.
The radial frequency is governed by the curvature of the effective
potential together with the non-trivial Schwarzschild kinetic factor,
whereas the vertical frequency receives an additional magnetic
contribution associated with the angular structure of the dipole
field. On the outward-Lorentz branch this contribution reduces the
vertical restoring frequency. Consequently,
$\Omega_\theta<\Omega_\phi$ wherever the orbit remains vertically
stable, and the vertical frequency vanishes at the vertical
stability boundary. In the strong-coupling regime the inner stable
orbit is therefore a vertically soft configuration rather than a
radially marginal one.

The different behaviour of the radial and vertical modes also gives
rise to two distinct patterns of epicyclic commensurabilities. When
radial marginal stability determines the inner boundary,
$\Omega_r/\Omega_\theta$ begins from zero and evolves through
sub-unity rational ratios. When vertical marginal stability determines
the boundary, the same ratio instead diverges near the inner orbit
and crosses super-unity commensurabilities before approaching the
weak-field regime at large radius. These crossings provide natural
candidate locations for nonlinear radial--vertical coupling, but the
linear epicyclic analysis alone does not establish resonance capture,
mode transfer, or chaotic dynamics.

The marginally bound circular orbit displays a different evolution
from the stability boundary. At weak coupling it is displaced inward
together with the radially marginal orbit. At stronger coupling,
however, its continuation follows a different algebraic circular-orbit
branch and approaches the near-horizon region rather than tracking
the outward-moving vertically limited stability boundary. The
marginally bound and marginally stable orbits therefore probe
different aspects of the charged-particle phase space once the
magnetic interaction becomes dominant.

An equally important conclusion is the strict separation between the
Lorentz-force dynamics of massive charged particles and the
null-geodesic structure of the Schwarzschild background. The
particle-dependent coupling $\beta$ has no analogue for neutral
photons. Since the electromagnetic field is treated as a test field
and does not alter the metric, the Schwarzschild photon sphere and
the associated vacuum critical impact parameter remain unchanged.
Magnetic fields can nevertheless have a strong observational impact
by modifying the dynamics, distribution, temperature, polarization,
and emissivity of the charged matter through which the photons
propagate.

The astrophysical interpretation of the charged-particle coupling
requires particular care. The parameter $\beta$ is not a property of
the compact object alone: it depends explicitly on the effective
specific charge of the orbiting particle or structure. Elementary
particles, ions, charged grains, and weakly charged coherent
structures may therefore explore very different dynamical regimes in
the same external magnetic field. A large value of $|\beta|$ should
also not be confused with significant electromagnetic backreaction on
the geometry, since the former is strongly influenced by the
particle charge-to-mass ratio.

For neutron stars, the stellar surface provides an additional
physical inner boundary and can hide part of the formal
charged-particle orbital structure. For black holes, the dipole field
must instead be externally supported by surrounding currents and
should not be interpreted as an intrinsic Schwarzschild magnetic
moment. In either case, the test-particle stability boundary should
not be identified automatically with the inner edge of a
quasi-neutral accretion flow, whose dynamics is collective and
requires a plasma or relativistic magnetohydrodynamic description.

The present results therefore provide a natural bridge toward more
realistic applications. One immediate extension is to replace the
leading asymptotic dipole approximation by the full relativistic
exterior field in the region where that solution is physically
appropriate, thereby quantifying the electromagnetic curvature
corrections to the orbital structure. A second direction is the
nonlinear dynamics of the radial and vertical modes, for which the
commensurabilities found here suggest a systematic study using
higher-order perturbation theory, Poincar\'e sections, frequency-map
analysis, and Lyapunov diagnostics.

Dissipative effects provide another natural development. Radiation
reaction can drive particles slowly through the circular-orbit and
commensurability structure derived here, potentially producing
capture, mode excitation, secular drift, or transitions between
different regions of phase space. The conservative solutions obtained
in the present work therefore provide the reference trajectories
needed for such an analysis, including the radiation-reaction
treatment pursued in the companion work~\cite{CRSP_react}.

More broadly, quantitative astrophysical applications will require
embedding the single-particle dynamics in a self-consistent
magnetospheric environment. This includes the effects of electric
fields, plasma screening, collective stresses, collisions, turbulent
transport, and radiative transfer. For horizon-scale observations in
particular, the distinction between charged-particle dynamics and
photon propagation suggests a useful separation of tasks: the
magnetic field determines how the emitting matter is organized and
radiates, whereas the background geometry determines the vacuum
photon trajectories.

The Schwarzschild--dipole system thus provides a useful relativistic
extension of the classical St{\o}rmer problem in which gravity,
electromagnetism, and orbital stability can be disentangled
analytically. The principal new feature of the outward-Lorentz branch
is the competition between radial and vertical stability, leading to
a transition from a radially limited weak-coupling regime to a
vertically limited strong-coupling regime. This structure provides a
well-defined starting point for the successive inclusion of nonlinear
dynamics, dissipation, realistic magnetospheres, collective plasma
physics, and observable emission.

\acknowledgments{
	FSNL acknowledges funding from the Funda\c{c}\~{a}o para a Ci\^{e}ncia e a Tecnologia (FCT) through national funds under the research grant UID/04434/2025 (DOI 10.54499/UID/04434/2025), and support from the FCT Scientific Employment Stimulus contract with reference CEECINST/00032/2018.}

\appendix
\section{The opposite Lorentz-force branch: $\beta<0$}
\label{app:negative_beta}

The main text has focused on positive azimuthal motion on the branch
$\beta>0$, for which the radial Lorentz force is directed outward.
The complementary orientation is obtained by writing
\begin{equation}
	\beta=-b,
	\qquad
	b=|\beta|>0 ,
	\label{eq:negative_beta_def}
\end{equation}
so that the Lorentz force is directed inward for positive azimuthal
motion. This branch is worth considering separately because changing
the relative orientation of the electromagnetic coupling and the
azimuthal motion alters not only the radial force balance but also
the vertical stability, the strong-coupling asymptotics, and the
structure of the marginally bound sequence.

The general dimensionless mechanical angular momentum is
$j=\lambda+\beta/x$. Hence, on the present branch,
\begin{equation}
	j=\lambda-\frac{b}{x}.
	\label{eq:j_negative_beta}
\end{equation}
We restrict attention to the positive-mechanical-angular-momentum
family $j>0$, continuously connected to the positive-angular-momentum
Schwarzschild circular-orbit sequence as $b\rightarrow0$.

It is useful to emphasize that the direction of the radial Lorentz
force is determined by the relative sign of the magnetic coupling
and the azimuthal motion. Since $\dot{\phi}$ has the sign of $j$, the
relevant combination is $\beta j$. The effective potential is also
invariant under the simultaneous transformation
$(\lambda,\beta,j)\rightarrow(-\lambda,-\beta,-j)$. Consequently,
the $\beta<0$, $j>0$ family considered here is related by reversal
of the azimuthal orientation to the $\beta>0$, $j<0$ family.
Throughout this Appendix we fix $j>0$, so that $\beta<0$
unambiguously corresponds to the inward-Lorentz orientation.

\subsection{Circular-orbit branch and force balance}
\label{app:negative_beta_circular}

The circular-orbit solution follows directly from
Eq.~\eqref{eq:lambda_circ} after setting $\beta=-b$. To display the
force balance more transparently, define
\[
D(x)=4x^2-8x+1 .
\]
On a circular orbit, the condition
$\partial_xV_{\rm eff}=0$ may be written directly in terms of the
mechanical angular momentum as
\begin{equation}
	(2x+1)^4
	-4xD(x)j^2
	+4b(4x^2-1)j
	=0 .
	\label{eq:negative_beta_circular_j}
\end{equation}
For the Schwarzschild-connected positive-$j$ solution,
\begin{equation}
	\begin{split}
		j_{\mathrm{in}}(x,b)
		&= \frac{2x+1}{2xD(x)}
		\left[ \sqrt{ b^2(2x-1)^2 + x(2x+1)^2D(x) } \right. \\
		&\quad \left. + b(2x-1) \right],
		\label{eq:negative_beta_j_exact}
	\end{split}
\end{equation}
The corresponding canonical angular momentum is
$\lambda=j_{\rm in}+b/x$.

For $x>x_{\rm ph}$, where $D(x)>0$,
Eq.~\eqref{eq:negative_beta_j_exact} shows explicitly that
$j_{\rm in}>0$ and that the mechanical angular momentum increases
with $b$ at fixed radius. This is the expected force-balance
response: gravity and the Lorentz force are both directed inward, so
a larger centrifugal contribution is required to maintain circular
motion. Since $\lambda=j+b/x$, the canonical angular momentum is
larger still.

The circular-orbit energy and orbital frequency retain the forms
derived in the main text,
$\mathcal{E}^2=\Phi^2[1+j^2/(\Psi^4x^2)]$ and
$M\Omega_\phi=\Phi^2j/(\mathcal{E}\Psi^4x^2)$. Consequently, at a
fixed radius the inward-Lorentz branch requires a larger mechanical
angular momentum than the corresponding neutral Schwarzschild orbit
and has a correspondingly larger specific energy and orbital
frequency. This behaviour is the opposite of that found on the
outward-Lorentz branch.

\subsection{Circular-orbit stability}
\label{app:negative_beta_stability}

Radial marginal stability is again determined by
$\partial_xV_{\rm eff}=0$ and
$\partial_x^2V_{\rm eff}=0$ at fixed $(\lambda,\beta)$. Eliminating
$\lambda$ gives exactly the same polynomial in $x$ and $|\beta|$ as
in the main text,
\begin{align}
	0={}&
	(2x-1)^2(2x+1)^4
	(4x^2-20x+1)^2
	\nonumber\\
	&-32b^2(2x+1)^2
	\left(
	80x^5-272x^4+368x^3
	\right.
	\nonumber\\
	&\hspace{2.4cm}
	\left.
	-224x^2+11x+1
	\right)
	\nonumber\\
	&-128b^4(2x-1)^2
	(4x^2-8x+7).
	\label{eq:negative_beta_radial_polynomial}
\end{align}
The fact that this equation contains only even powers of the coupling
does \emph{not} imply identical stability boundaries for the two
Lorentz-force orientations. Eliminating $\lambda$ removes the sign
information needed to identify the relevant circular-orbit branch.
For $\beta<0$, the physical root must be evaluated on
$\lambda=\lambda_+(x,-b)$ and connected continuously to the
positive-angular-momentum Schwarzschild family.

At weak coupling the radial stability boundary moves outward:
\begin{equation}
	x_{\rm rISCO}
	=
	x_{\rm ISCO}^{(0)}
	+
	\left(
	\frac{3\sqrt3}{8}
	+\frac{\sqrt2}{4}
	\right)b
	+\mathcal{O}(b^2).
	\label{eq:negative_beta_weak_isco}
\end{equation}
This is the exact sign reversal of the leading displacement obtained
for the outward-Lorentz branch.

The vertical dynamics changes even more directly. The general
frequency relation becomes
\begin{equation}
	\frac{\Omega_\theta^2}{\Omega_\phi^2}
	=
	1+\frac{2b}{jx}.
	\label{eq:negative_beta_vertical}
\end{equation}
Hence $\Omega_\theta>\Omega_\phi$ everywhere on the positive-$j$
family. The magnetic interaction enhances the vertical restoring
frequency relative to the orbital frequency, rather than suppressing
it, and $\Omega_\theta^2$ cannot vanish for $b>0$ and $j>0$.
There is therefore no analogue of the vertical marginal-stability
boundary or of the critical coupling $\beta_{\rm c}$ encountered on
the outward-Lorentz branch. The inner boundary of fully stable
circular motion remains radially determined.

This distinction is particularly important at strong coupling.
Seeking a large-$b$ radial marginal orbit in the form
$x=C_-b^{2/3}$ and retaining the leading terms of
Eq.~\eqref{eq:negative_beta_radial_polynomial} gives
\[
C_-^6-10C_-^3-2=0 .
\]
The positive solution is $C_-^3=5+3\sqrt3$, so that
\begin{equation}
	x_{\rm ISCO}
	\sim
	(5+3\sqrt3)^{1/3}b^{2/3},
	\qquad
	b\gg1 .
	\label{eq:negative_beta_strong_isco}
\end{equation}
Thus the large-radius root that does not determine the physical
strong-coupling boundary on the $\beta>0$
Schwarzschild-connected family becomes precisely the relevant radial
stability boundary on the inward-Lorentz branch.

The associated strong-coupling orbital quantities can also be
obtained analytically. Writing
$C_-=(5+3\sqrt3)^{1/3}$, one finds
\begin{align}
	j_{\rm ISCO}
	&\sim
	\frac{2+\sqrt3}{C_-}\,b^{1/3},
	&
	\lambda_{\rm ISCO}
	&\sim
	\frac{3+\sqrt3}{C_-}\,b^{1/3},
	\nonumber\\
	M\Omega_{\phi,\rm ISCO}
	&\sim
	\frac{1}{(1+\sqrt3)b},
	&
	\frac{\Omega_{\theta,\rm ISCO}}
	{\Omega_{\phi,\rm ISCO}}
	&\longrightarrow
	\sqrt{5-2\sqrt3}.
	\label{eq:negative_beta_strong_quantities}
\end{align}
Numerically,
$\sqrt{5-2\sqrt3}\simeq1.23931$, so the
vertical-to-orbital frequency ratio remains finite and larger than
unity in the strong-coupling limit. Both frequencies themselves
decrease as $b^{-1}$ along the outward-moving stability boundary.
The radial frequency, by contrast, vanishes at the ISCO by
definition.

The specific energy tends to unity from below as
$b\rightarrow\infty$; more explicitly,
$\mathcal{E}_{\rm ISCO}
=1-(3-\sqrt3)(4C_-)^{-1}b^{-2/3}
+\mathcal{O}(b^{-4/3})$. Although the gravitational potential becomes
weak because $x_{\rm ISCO}\rightarrow\infty$, gravity does not
become dynamically irrelevant. On the scaling
$x\propto b^{2/3}$, the gravitational, centrifugal, and magnetic
contributions to the radial force remain of the same asymptotic
order.

\subsection{Relation to the flat-space St{\o}rmer limit}
\label{app:negative_beta_stormer}

The inward-Lorentz orientation is also the one that admits a
non-trivial equatorial magnetic circular orbit in an auxiliary
gravity-free comparison. Formally setting $\Phi=\Psi=1$ while
retaining the dimensionless variables and leading dipole interaction
used above gives
\begin{equation}
	V_{\rm flat}
	=
	1+
	\frac{1}{x^2}
	\left(
	\lambda-\frac{b}{x}
	\right)^2 ,
	\;
	\frac{dV_{\rm flat}}{dx}
	=
	-\frac{2(\lambda x-b)(\lambda x-2b)}{x^5}.
	\label{eq:negative_beta_flat}
\end{equation}
This construction is an auxiliary gravity-free limit of the
dimensionless dynamical equations and should not be confused with
taking a literal $M\rightarrow0$ limit while keeping the definitions
of $x$ and $b$ fixed.

Besides the trivial zero-mechanical-angular-momentum solution
$\lambda x=b$, the moving circular orbit satisfies
$\lambda x=2b$, or equivalently $j=b/x$. The inward magnetic force
then supplies the required centripetal acceleration.

This flat-space orbit is, however, radially unstable: evaluated on
the non-trivial circular solution,
$d^2V_{\rm flat}/dx^2=-4b^2/x^6<0$. Its vertical frequency is
enhanced relative to the orbital one, with
$\Omega_\theta/\Omega_\phi=\sqrt3$ in the same auxiliary limit.

This observation clarifies the physical content of the
strong-coupling Schwarzschild result. The orbit
\eqref{eq:negative_beta_strong_isco} is not obtained simply by
discarding gravity and retaining a magnetically confined flat-space
St{\o}rmer orbit. Even though the ISCO moves to a weak-field region,
gravity remains essential to the radial stability balance. The
large-$b$ regime is therefore a magnetically modified weak-gravity
limit rather than a purely magnetic one.

\subsection{Marginally bound circular orbit}
\label{app:negative_beta_mb}

The marginally bound orbit again satisfies $V_{\rm eff}=1$ together
with $\partial_xV_{\rm eff}=0$ at fixed canonical angular momentum.
The energy condition gives the same positive mechanical angular
momentum as in the main text, while the relation
$j=\lambda-b/x$ changes the circularity condition to
\begin{equation}
	\sqrt{x_{\rm mb}}\,(2x_{\rm mb}+1)
	\left(4x_{\rm mb}^2-12x_{\rm mb}+1\right)
	=
	2\sqrt{2}\,b\,(2x_{\rm mb}-1)^2 .
	\label{eq:negative_beta_mb}
\end{equation}

An immediate consequence of the sign in
Eq.~\eqref{eq:negative_beta_mb} is that every exterior solution with
$b>0$ satisfies $x_{\rm mb}>x_{\rm mb}^{(0)}$. Indeed, the
right-hand side is positive, so in the exterior domain $x>1/2$ the
factor $4x^2-12x+1$ on the left must be positive; on the exterior
Schwarzschild-connected branch this requires
$x>(3+2\sqrt2)/2$. Thus the outward displacement of the marginally
bound orbit is not merely a weak-coupling trend but persists
throughout the positive-$j$ continuation.

Expanding for small $b$ gives
\begin{equation}
	x_{\rm mb}
	=
	x_{\rm mb}^{(0)}
	+\frac{b}{2}
	+\mathcal{O}(b^2).
	\label{eq:negative_beta_mb_weak}
\end{equation}
Unlike the $\beta>0$ case, no algebraic-branch coalescence occurs
along this marginally bound sequence. Since
$x_{\rm mb}>x_{\rm mb}^{(0)}>x_{\rm ph}$, one has $D(x)>0$, and the
discriminant entering Eq.~\eqref{eq:lambda_circ} remains strictly
positive. The marginally bound orbit therefore stays on the
Schwarzschild-connected $\lambda_+(x,-b)$ branch for all $b>0$.

For strong coupling, balancing the leading terms of
Eq.~\eqref{eq:negative_beta_mb} gives
$x_{\rm mb}^{3/2}\simeq\sqrt2\,b$, and hence
\begin{equation}
	x_{\rm mb}
	\sim
	2^{1/3}b^{2/3},
	\qquad
	b\gg1 .
	\label{eq:negative_beta_mb_strong}
\end{equation}
The corresponding mechanical and canonical angular momenta behave as
$j_{\rm mb}\sim2^{2/3}b^{1/3}$ and
$\lambda_{\rm mb}\sim3\,2^{-1/3}b^{1/3}$, while
$M\Omega_{\phi,\rm mb}\sim b^{-1}$. The
vertical-to-orbital frequency ratio tends to $\sqrt2$.

Both the marginally bound and marginally stable radii therefore move
outward as $b^{2/3}$, although with different coefficients. Their
strong-coupling ratio tends to
\[
\frac{x_{\rm mb}}{x_{\rm ISCO}}
\longrightarrow
\left(
\frac{2}{5+3\sqrt3}
\right)^{1/3}
\simeq0.581 ,
\]
so the marginally bound orbit remains well inside the radial
stability boundary.

This behaviour also has an important consequence for the
electromagnetic approximation. On the inward-Lorentz branch both
strong-coupling radii migrate toward $x\gg1$, where the leading
asymptotic dipole potential becomes locally better controlled. This
contrasts sharply with the formal strong-coupling marginally bound
continuation on the outward-Lorentz branch, which approaches the
near-horizon region where the leading-dipole approximation is no
longer quantitatively reliable.

\subsection{Epicyclic hierarchy and frequency commensurabilities}
\label{app:negative_beta_frequencies}

The absence of vertical marginality leads to a frequency structure
qualitatively different from the strong-coupling behaviour discussed
in the main text. At the inner boundary of the stable sequence,
$\Omega_r=0$ while $\Omega_\theta$ remains finite and satisfies
$\Omega_\theta>\Omega_\phi$. Hence
$\Omega_r/\Omega_\theta\rightarrow0$ as the ISCO is approached from
above for every value of $b$ on the Schwarzschild-connected
positive-$j$ branch.

At fixed $b$ and sufficiently large radius, the Schwarzschild
weak-field regime is recovered. Expanding the exact circular-orbit
frequencies along the physical branch gives
\begin{equation}
	\frac{\Omega_r^2}{\Omega_\theta^2}
	=
	1-\frac{6}{x}
	-\frac{6b}{x^{3/2}}
	+\frac{6}{x^2}
	+\mathcal{O}(x^{-5/2}),
	\qquad
	x\rightarrow\infty .
	\label{eq:negative_beta_frequency_asymptotic}
\end{equation}
Thus $\Omega_r/\Omega_\theta$ approaches unity from below
asymptotically. Similarly,
$\Omega_\theta^2/\Omega_\phi^2
=1+2b/x^{3/2}+\mathcal{O}(x^{-5/2})$, showing explicitly that the
magnetic enhancement of the vertical frequency relative to the
orbital frequency disappears at large radius.

There is therefore no strong-coupling vertical pole analogous to
$\Omega_r/\Omega_\theta\rightarrow\infty$ on the outward-Lorentz
branch. The natural low-order radial--vertical commensurabilities on
the stable inward-Lorentz sequence are instead associated with a
ratio that starts from zero at radial marginal stability and
approaches unity in the weak-field region. As in the main text, such
rational frequency crossings are only kinematical indicators; a
genuine resonance requires nonlinear coupling between the two
degrees of freedom.

\subsection{Comparison of the two Lorentz-force orientations}
\label{app:beta_comparison}

The principal differences between the two relative Lorentz-force
orientations are summarized in
Table~\ref{tab:beta_branch_comparison}. The comparison assumes
positive azimuthal motion, $j>0$, so that the sign of $\beta$
directly specifies whether the Lorentz force is outward or inward.

\begin{table*}[h!]
	\centering
	\caption{Comparison of the outward- and inward-Lorentz
		orientations for positive azimuthal motion and positive
		mechanical angular momentum. On the negative-coupling branch,
		$b=-\beta=|\beta|>0$.}
	\label{tab:beta_branch_comparison}
	
	\renewcommand{\arraystretch}{1.45}
	\setlength{\tabcolsep}{8pt}
	
	\begin{ruledtabular}
		\begin{tabular}{|c|c|c|} 
			
			\textbf{Property}
			&
			\textbf{$\boldsymbol{\beta>0}$: outward Lorentz force}
			&
			\textbf{$\boldsymbol{\beta<0}$: inward Lorentz force}
			\\
			
			\hline
			
			\parbox[t]{0.22\textwidth}{\raggedright Mechanical angular momentum}
			&
			\parbox[t]{0.34\textwidth}{\raggedright $j=\lambda+\beta/x$; outward electromagnetic support reduces the mechanical angular momentum required at fixed radius.}
			&
			\parbox[t]{0.34\textwidth}{\raggedright $j=\lambda-b/x$; inward electromagnetic attraction increases the mechanical angular momentum required at fixed radius.}
			\\
			\hline
			\parbox[t]{0.22\textwidth}{\raggedright Weak-coupling radial boundary}
			&
			\parbox[t]{0.34\textwidth}{\raggedright Shifted inward relative to Schwarzschild.}
			&
			\parbox[t]{0.34\textwidth}{\raggedright Shifted outward relative to Schwarzschild.}
			\\
			\hline
			\parbox[t]{0.22\textwidth}{\raggedright Relative vertical restoring frequency}
			&
			\parbox[t]{0.34\textwidth}{\raggedright Reduced:\\ $\Omega_\theta<\Omega_\phi$ on vertically stable orbits.}
			&
			\parbox[t]{0.34\textwidth}{\raggedright Enhanced:\\ $\Omega_\theta>\Omega_\phi$ on the positive-$j$ family.}
			\\
			\hline
			\parbox[t]{0.22\textwidth}{\raggedright Vertical marginal stability}
			&
			\parbox[t]{0.34\textwidth}{\raggedright Appears at sufficiently strong coupling and eventually controls the inner fully stable orbit.}
			&
			\parbox[t]{0.34\textwidth}{\raggedright Absent on the Schwarzschild-connected positive-$j$ family.}
			\\
			\hline
			\parbox[t]{0.22\textwidth}{\raggedright Inner fully stable boundary}
			&
			\parbox[t]{0.34\textwidth}{\raggedright Radially limited at weak coupling and vertically limited at strong coupling.}
			&
			\parbox[t]{0.34\textwidth}{\raggedright Radially limited for all couplings on the branch considered.}
			\\
			\hline
			\parbox[t]{0.22\textwidth}{\raggedright Strong-coupling stability radius}
			&
			\parbox[t]{0.34\textwidth}{\raggedright $x_{\rm ISCO}\sim6^{1/3}\beta^{2/3}$, set by vertical marginality.}
			&
			\parbox[t]{0.34\textwidth}{\raggedright $x_{\rm ISCO}\sim (5+3\sqrt3)^{1/3}b^{2/3}$, set by radial marginality.}
			\\
			\hline
			\parbox[t]{0.22\textwidth}{\raggedright Strong-coupling orbital frequency}
			&
			\parbox[t]{0.34\textwidth}{\raggedright $M\Omega_{\phi,\rm ISCO}\sim(3\beta)^{-1}$.}
			&
			\parbox[t]{0.34\textwidth}{\raggedright $M\Omega_{\phi,\rm ISCO} \sim[(1+\sqrt3)b]^{-1}$.}
			\\
			\hline
			\parbox[t]{0.22\textwidth}{\raggedright Weak-coupling marginally bound orbit}
			&
			\parbox[t]{0.34\textwidth}{\raggedright Shifted inward.}
			&
			\parbox[t]{0.34\textwidth}{\raggedright Shifted outward.}
			\\
			\hline
			\parbox[t]{0.22\textwidth}{\raggedright Strong-coupling marginally bound orbit}
			&
			\parbox[t]{0.34\textwidth}{\raggedright Changes algebraic branch and formally approaches the isotropic horizon; the leading-dipole approximation then ceases to be controlled.}
			&
			\parbox[t]{0.34\textwidth}{\raggedright Remains on the Schwarzschild-connected branch and moves outward as $x_{\rm mb}\sim2^{1/3}b^{2/3}$.}
			\\
			\hline
			\parbox[t]{0.22\textwidth}{\raggedright Flat-space magnetic circular orbit}
			&
			\parbox[t]{0.34\textwidth}{\raggedright Positive azimuthal motion has the wrong Lorentz-force orientation for magnetic centripetal confinement.}
			&
			\parbox[t]{0.34\textwidth}{\raggedright A non-trivial magnetic circular orbit exists in the auxiliary gravity-free problem, but is radially unstable.}
			\\
			\hline
			\parbox[t]{0.22\textwidth}{\raggedright Radial--vertical frequency ratio}
			&
			\parbox[t]{0.34\textwidth}{\raggedright Can diverge at a vertically marginal inner boundary and pass through super-unity commensurabilities.}
			&
			\parbox[t]{0.34\textwidth}{\raggedright Starts from zero at the radially marginal inner boundary; there is no vertical-soft divergence, and the ratio tends to unity from below asymptotically.}
			\\
			\hline
			\parbox[t]{0.22\textwidth}{\raggedright Strong-coupling validity of the leading dipole term}
			&
			\parbox[t]{0.34\textwidth}{\raggedright The fully stable boundary moves to large radius, but the formal marginally bound continuation approaches the near-horizon region.}
			&
			\parbox[t]{0.34\textwidth}{\raggedright Both the stability and marginally bound scales move toward $x\gg1$, improving the local asymptotic control of the dipole approximation.}
			\\
			
		\end{tabular}
	\end{ruledtabular}
	
	\renewcommand{\arraystretch}{1}
\end{table*}

The comparison shows that reversing the relative orientation of the
Lorentz force is not a trivial relabelling of the same circular-orbit
sequence. Although both branches recover Schwarzschild dynamics as
$|\beta|\rightarrow0$, their finite-coupling stability structures
differ fundamentally. Outward Lorentz support weakens the vertical
restoring force and ultimately produces a vertically limited
strong-coupling regime. Inward Lorentz forcing instead enhances the
vertical restoring frequency relative to the orbital motion, leaving
radial marginality as the mechanism that determines the inner fully
stable orbit.

The energetic structure differs just as strongly. On the
inward-Lorentz branch, the marginally bound and marginally stable
orbits remain on the Schwarzschild-connected family and both move
outward at strong coupling. On the outward-Lorentz branch, the fully
stable boundary also moves outward once vertical marginality takes
over, but the marginally bound solution changes algebraic branch and
formally evolves toward the horizon. The relative sign of the
Lorentz force therefore governs not only the local radial force
balance but also the global organization of the charged-particle
circular-orbit phase space.

\newpage


\end{document}